\documentclass[a4paper,11pt]{article}
\pdfoutput=1 

\usepackage{jheppub} 

\usepackage[T1]{fontenc} 
\usepackage[export]{adjustbox}
\usepackage{amssymb,amscd,braket}

\title{On Local and Integrated Stress-Tensor Commutators}

\author[1]{Mert Be\c sken}
\author[1]{Jan de Boer,}
\author[1]{Gr\'egoire Mathys}

\affiliation[1]{Institute for Theoretical Physics and Delta Institute for Theoretical Physics,\\University of Amsterdam, PO Box 94485, 1090 GL Amsterdam, The Netherlands}

\emailAdd{m.besken@uva.nl}
\emailAdd{j.deboer@uva.nl}
\emailAdd{g.o.mathys@uva.nl}

\abstract{We discuss some general aspects of commutators of local operators in Lorentzian CFTs, which can be obtained from a suitable analytic
	continuation of the Euclidean operator product expansion (OPE). Commutators only make sense as distributions, and care has to be taken to 
	extract the right distribution from the OPE. We provide explicit computations in two and four-dimensional CFTs, focusing mainly on
	commutators of components of the stress-tensor. We rederive several familiar results, such as the 
	canonical commutation relations of free field theory, the local form of the Poincar\'e algebra, and the Virasoro algebra of two-dimensional CFT. 
	We then consider commutators of light-ray operators built from the stress-tensor. Using simplifying features of the light sheet limit in four-dimensional CFT 
	we provide a direct computation of the BMS algebra formed by a specific set of light-ray operators in theories with no light scalar conformal primaries. 
	In four-dimensional CFT we define a new infinite set of light-ray operators constructed from the stress-tensor, which all have well-defined matrix
	elements. These are a direct generalization of the two-dimensional Virasoro light-ray operators that are obtained from a conformal 
	embedding of Minkowski space in the Lorentzian cylinder. They obey Hermiticity conditions similar to their two-dimensional analogues, 
	and also share the property that a semi-infinite subset annihilates the vacuum. 
}

\def\a{\alpha}
\def\b{\beta}
\def\g{\gamma}
\def\G{\Gamma}
\def\eps{\epsilon}

\def\om{\omega}
\def\t{\tau}
\def\m{\mu}
\def\n{\nu}
\def\s{\sigma}
\def\r{\rho}
\def\D{\Delta}
\def\d{\delta}

\def\r{\rho}

\def\cA{{\cal A}}

\def\cC{{\cal C}}

\def\cE{{\cal E}}

\def\cI{{\cal I}}

\def\cK{{\cal K}}
\def\cL{{\cal L}}

\def\cN{{\cal N}}
\def\cO{{\cal O}}

\DeclareSymbolFont{matha}{OML}{txmi}{m}{it}
\DeclareMathSymbol{\vv}{\mathord}{matha}{118}

\newcommand{\p}{\partial}
\newcommand {\be} {\begin {equation}}
\newcommand {\ee} {\end {equation}}

\newcommand{\no}{\nonumber}
\newcommand\zb{\bar{z}}
\newcommand{\<}{\langle}
\renewcommand{\>}{\rangle}
\newcommand\TTten{(TT)}
\newcommand\TTTten{(TTT)}

\begin{document} 
\maketitle
\flushbottom

\section{Introduction}

Commutators of local operators belong to the fundamental building blocks of any Lorentzian quantum field theory. They are the starting point
of canonical quantization, diagnose causality and locality, and can be used to find sum rules and dispersion relations. Of particular interest
are equal-time commutators such as those between the fundamental fields and their conjugate momenta. For free fields one can for example
use the field equations to construct general unequal time commutators starting from the equal-time commutators. However, in a general interacting
quantum field theory it is much more difficult to study equal time commutators. If a perturbative Lagrangian description is available, one
can try to construct equal-time commutators of composite operators in perturbation theory, but this procedure is plagued by normal ordering
ambiguities and UV divergences. If the operators correspond to the components of a conserved current associated to a global symmetry, their
equal-time commutators need to take a particular form in order to be consistent with current conservation and the global symmetries \cite{SchwingerNonAbelian,SchwingerCommutationConservation,BoulwareDeser,Trubatch}. The current
algebra program which was initiated in the 60's \cite{Gell-Mann,Sugawara,Sommerfield} tried to exploit equal-time commutators of currents as much as possible, 
leading to e.g. the Adler-Weisberger formula, and could even be used in cases with weakly broken global symmetries such as partially 
conserved axial currents \cite{CurrentAlgebraandPCAC}. 

One important issue in defining equal-time commutators is that in Lorentzian quantum field theory only smeared operators are well-defined, 
and this smearing needs to be done both in space and time. It is therefore a priori unclear whether an unambiguous notion of equal-time commutators
exists, even for conserved currents, in a distributional sense (for a discussion of various issues in taking the equal-time limit
see e.g. \cite{Brandt}). We will return to this issue below.

In this paper we study commutators of local and integrated current components in conformal field theories in two and four dimensions, focusing on
components of the stress-tensor. The techniques that we develop should however also be applicable for other conserved currents in CFTs in $d\neq 2,4$. The main
motivation for this work is to compute, as much as possible from first principles, commutators of various types of light-ray operators 
built out of components of the stress-tensor. Such operators are very interesting, as one can use them to build a BMS algebra \cite{Cordova:2018ygx,Donnay_2020}, to 
construct the Averaged Null Energy Operator (ANEC) \cite{Faulkner:2016mzt,Hartman:2016lgu} (see also \cite{Hofman:2008ar,Hofman:2016awc,Cordova:2017zej,Meltzer:2018tnm,Belin:2019mnx,Kologlu:2019mfz,Kologlu:2019bco,Chang:2020qpj} for example of other uses of the ANEC operator in CFT), and to possibly uncover new symmetry-like structures generated by operators of the form
$\int du f(u) T_{uu}$ with $u$ a lightcone coordinate \cite{Casini:2017vbe,Cordova:2018ygx,Huang:2019fog,Huang:2020ycs,belin2020stress,Huang:2021hye}. For a general discussion of light-ray operators in CFT, see e.g. \cite{Kravchuk:2018htv}. In addition, recent evidence \cite{Fitzpatrick:2019zqz,Huang:2019fog,Kulaxizi:2019tkd,Fitzpatrick:2019efk,Li:2019tpf,Karlsson:2019dbd,Li:2019zba,Karlsson:2020ghx,Li:2020dqm,Karlsson_2020,Fitzpatrick:2020yjb} suggests a certain universality in the multi-stress-tensor sector in a class of $d>2$ CFTs. In $d=2$, such universality is a consequence of Virasoro symmetry. This motivates the search for a Virasoro-like structure in higher-dimensional CFTs. To study commutators of light-ray operators we will need to consider general commutators, not just equal-time commutators, which in general have
support when points are null- or time-like separated. 

To properly define commutators, we will start with the Euclidean expression for the product of two operators (defined through the 
operator product expansion) and then analytically continue to Lorentzian signature. To obtain an ordered product of Lorentzian operators
of the form $A_1(t_1,x_1)\ldots A_p(t_p,x_p)$ we need to give the time variables a small imaginary part such that ${\rm Im}(t_1)<\ldots
<{\rm Im} (t_p)$. These small imaginary parts allow for an unambiguous analytic continuation and are nothing but a version of the usual
$i\epsilon$ prescription of quantum field theory (for a nice review, see \cite{Hartman:2015lfa}). This is also how one infers that operators which are smeared in space and time have 
well-defined correlation functions: one first computes everything for small but finite imaginary contributions to the time coordinates, and
takes contributions to zero at the end of the computation. The same procedure can also be followed to study equal-time commutators. 
Here one may or may not be left with a well-defined $\epsilon\rightarrow 0$ limit. If the limit exists and gives rise to a well-defined
spatial distribution (like e.g. $\delta(\vec{x}-\vec{y})$) we will call the equal-time commutator well-defined. In general there will also
be terms which do not have a well-defined $\epsilon\rightarrow 0$ limit which represent UV-divergences in the equal-time commutators. This 
procedure gives rise to an unambiguous notion of equal-time commutator in cases in which the equal-time commutator is well-defined, and we will
illustrate this procedure with several examples below.

To find the c-number (or Schwinger) terms in commutators \cite{SchwingerFieldCommutators,BoulwareDeser,Brandt,Mahanthappa:1969bg}, it is sufficient to know the c-number contributions to the OPE of two operators, 
which is equal to the vacuum two-point function of the two operators. To find contributions of other operators, we do in principle need the full OPE.
The terms in the OPE which only contain singularities of the form $(x_1-x_2)^{-k}$ with $k$ an even integer give rise to 
contributions which only have support on the lightcone, as these terms are analytic functions of time. For other values of $k$, which are generically present in OPEs, there are other contributions which have support not just on but also inside the lightcone. Examining the full
operator content of commutators is beyond the scope of this work, and we will mostly look at specific examples only. We will also find it
more convenient to extract the contribution of an operator ${\cal O}$ to a commutator $[A,B]$ by examining three-point functions of the form
$\langle [A,B]{\cal O}\rangle $ rather then using the OPE directly. The drawback of using the OPE is that we need to keep track of
an operator and all its global conformal descendants for which in general no simple explicit expressions are available. The three-point functions
on the other hand are in general explicitly known, as they are fixed up to some numbers by conformal symmetry. 

We will use these techniques to revisit the construction of a BMS algebra from light-ray operators of the form $\int du\, g(u) T_{u\m}$ 
where $g(u)$ is a first order polynomial in $u.$ This algebra involves commutators with both light-ray operators placed on a 
common light sheet. After explaining our technique we first compute local stress-tensor commutators on a 
light sheet $\lim_{\vv \to 0}[T_{\m\n}(u,\vv,x_\perp),T_{\a\b}(0)]$ and argue that in a generic interacting CFT with no light scalars with dimension $1 \leq \D \leq 2$ in $d=4$ only the identity operator and stress-tensor contribute in the limit; this is due to the fact a light sheet contains only null or spacelike separated points and the OPE becomes an expansion in twist in the lightcone limit. We then integrate this commutator to derive the BMS algebra proposed in \cite{Cordova:2018ygx}. Light scalars have a potential to spoil the BMS algebra. We discuss the situation in theories with $\cN=1$ superconformal symmetry and argue these operators are not present based on the results of \cite{Cordova:2016emh}. This establishes a derivation of the BMS algebra in these theories.

We also consider more general
operators of the form $\int du f(u) T_{uu}$. In $d$ dimensions, matrix elements of $T_{uu}$ scale as $1/u^{d+2}$ as $u\rightarrow \infty$\footnote{This is true for three-point functions of $T_{uu}$ with two primary operators.}
and therefore we need $f(u)/u^{d+1}\rightarrow 0$ as $u\rightarrow \infty$ in order to have well-defined operators of this type. One might
have thought for example in two dimensions the usual Virasoro algebra is recovered by taking $f(u)=u^n$, but these operators do not
have well-defined matrix elements in general states. As we will show, by embedding the plane on the Lorentzian cylinder, a more natural
definition of the 2d Virasoro generators is to take $f(u)\sim (iR+u)^{1-n} (iR-u)^{1+n}$ where $R$ denotes the size of the circle, which does give rise to a well-defined family of
operators that indeed form the Virasoro algebra. The obvious 4d generalization of these operators arises by taking $f(u)
\sim (iR+u)^{2-n} (iR-u)^{2+n}$. These operators close under the subgroup of the conformal group which preserves the null ray \cite{Braun:2003rp,belin2020stress}, 
and also have the property that they annihilate the vacuum on the left and/or on the right, but it is not yet clear whether these 
operators give rise to interesting Ward identities or sum rules.

This paper is organized as follows. In section \ref{sec:gen} we discuss a few general features of our setup and discuss the features of 
stress-tensor commutators which follow from Poincar\'e invariance. In section \ref{sec:fs} we illustrate some of the ideas in the case of a free scalar field. In section \ref{sec:stcft2} we consider two-dimensional local and smeared commutators and recover the Virasoro algebra using the smearing functions $f(u)$ alluded to above. In section \ref{STcomm4d} we study commutators of local operators
built from various components of the stress-tensor in four dimensions, while in section \ref{sec:BMS} we consider commutators of light-ray operators
constructed out of the stress-tensor. We conclude in section \ref{sec:conc} with some open questions and future prospects, and higlight the further analysis required to understand the algebras we consider. Several of the more technical details are contained in the appendices.  

\subsection*{Notation}

We denote spacetime points by $x=x^\m.$ In four dimensions, we denote Cartesian coordinates as  $x=(x^0,x_1,x_2,x_3)=(t,\vec{x})=(t,x_i)$ and $r^2=\sum_{i=1}^3 x_i^2.$ We denote lightcone coordinates $u=t-x_1,~\vv=t+x_1$ and write $x=(u,\vv,x_2,x_3)=(u,\vv,x_\perp)=(u,\vv,x_A).$ We use the mostly plus metric. In lightcone coordinates $ds^2=-du d\vv +dx_\perp^2.$ We use $x_\perp^2=x_2^2+x_3^2.$ We use the same conventions in two dimensions with transverse coordinates omitted.

\section{Generalities}
\label{sec:gen}

In this section we outline our setup, which we use to compute commutators of local operators from the operator product expansion. For general operators we analyze the structure of singularities in the OPE and discuss the qualitative features of commutators they give rise to. Explicating the difficulties involved in computing generic commutators, we outline simplifications that arise in the equal-time and light sheet limits. We end with a review of implications of Poincar\'e invariance for stress-tensor commutators.

\subsection{Commutators from the OPE}
\label{sec:gencom}
In CFT local operators are organized in the OPE \cite{simmonsduffin2016tasi}
\begin{align}
\label{ope}
\cO_{i}(x) \cO_{j}(0)=\sum_{k} C_{ijk}^{\a\b\ldots}(x,\p)O_{\a\b\ldots}^{k}(0)\, .
\end{align}
As described in the Introduction, we compute the commutator of two operators using the $i\eps$ prescription
\begin{align}
\label{cd}
[\cO_i(t,\vec{x}),\cO_j(0)] = \lim_{\eps \to 0} \big[\cO_i(t -i\eps,\vec{x}) \cO_j(0) - \cO_i(t +i\eps,\vec{x}) \cO_j(0)\big]\, .
\end{align}
If we focus on a particular term in the OPE of the form $\cO_k(0)/x^{2h}$, with $\cO_k$ some local operator, its contribution
to the commutator takes the form of $\cO_k(0)$ times
\begin{align} \label{jan2}
f_h(t,\vec{x}) \equiv \lim_{\eps \to 0} \left( \frac{1}{(-(t-i\epsilon)^2 + r^2)^h} - \frac{1}{(-(t+i\epsilon)^2 + r^2)^h} \right)\, .
\end{align}
To investigate $f_h(t,\vec{x})$ viewed as a distribution in $t$, we integrate it against an arbitrary test function. 
The first term of $f_h(t,\vec{x})$ has poles in the upper half of the complex plane, the second term has poles in the lower half of the complex plane.
To integrate it against a test function $g(t)$ it is therefore advantageous to Fourier transform to frequency space where modes $e^{i\omega t}$ with
positive omega are analytic and bounded in the upper half of the complex $t$-plane, and negative frequency modes are analytic and bounded 
in the lower half of the complex $t$-plane. Assuming $h$ is integer, we can close the contour in either the upper or lower half of the 
complex plane for each Fourier mode $e^{i\omega t}$ and evaluate the integral $\int_{-\infty}^\infty f_h(t,\vec{x}) g(t) dt$. The final answer can indeed
be rewritten as a distribution in $t$ and we find that
\begin{align} \label{jan1}
f_h(t,\vec{x}) = -\frac{2\pi i }{(h-1)!} \left[ (t-r)^{-h} \partial_t^{h-1} \delta(t+r) +
(t+r)^{-h} \partial_t^{h-1} \delta(t-r) \right]\, .
\end{align}
This answer can easily be generalized to more general structures which can appear in OPEs \eqref{ope} as long as we encounter only terms with singularities
and no branch cuts in the complex $t$-plane.

For $h$ non-integer, one sees from the explicit expression that it must vanish for $t^2<r^2$ as the difference of the two terms
only receives a contribution from the region $t^2\geq r^2$ where the branch cut structure of the two terms becomes relevant. Therefore, 
for non-integer $h$, $f_h(t,\vec{x})$ has support on and inside the lightcone. One can think of this function as defining an analytic continuation
of the distribution (\ref{jan1}) to non-integer $h$. Alternatively, one can see that for $h<0$ and not an integer, $f_h(t,\vec{x}) \sim
\sin(\pi h) (t^2-r^2)^{-h} (\theta(t-r) + \theta(-t-r))$, and this answer can by analytically continued to positive $h$ 
and/or one can make $h$ positive by differentiating this expression. We will not consider the case of non-integer $h$ in the remainder of this
work.

\subsection*{Equal-time commutators}

In the example above, the function $f_h(t,\vec{x})$ as defined in (\ref{jan2}) vanishes for $t=0$ and would therefore give rise to a 
vanishing equal-time commutator. To get something more interesting we consider instead a contribution to the OPE of the form
$t^p/x^{2h}$ so that
\begin{align} \label{jan3}
f_{h,p}(t,\vec{x}) \equiv \lim_{\eps \to 0} \left( \frac{(t-i\epsilon)^p}{(-(t-i\epsilon)^2 + r^2)^h} - 
\frac{(t+i\epsilon)^p}{(-(t+i\epsilon)^2 + r^2)^h} \right)\, .
\end{align}
We can easily analyze \eqref{jan3} for $t=0$, thanks to the appropriate $i\epsilon$ prescription. It vanishes for $p$ even, and for odd $p$ we can examine it by integrating it
over $\vec{x}$. This shows that the expression is UV divergent for $p<2h+1-d$ and the equal-time commutator is ill-defined, 
while for $p=2h+1-d$ we obtain
\be
f_{h,p}(0,\vec{x}) = -2 i^p \frac{\Gamma(\frac{p}{2})}{\Gamma(h)} \pi^{\frac{d-1}{2}} \delta(\vec{x})\, ,
\ee
and for $p>2h+1-d$ we obtain zero. In principle the latter could also indicate that this case corresponds to derivatives of delta functions, 
but a simple scaling argument shows that that is impossible. Derivatives of delta functions correspond to linear combinations of
divergent $f_{h,p}$ such that the divergences cancel each other, as we will see in many examples later in the paper.

\subsection*{Light sheet commutators}

Exactly the same logic can be followed to study commutators in other kinematic regimes, such as a light sheet limit, 
in which $\vv \to 0$ where $\vv=t+x_1$ denotes a lightcone coordinate. 
This places the two operators on a common light sheet which is composed of non-timelike separated points. Non-vanishing 
contributions to the commutator on a light sheet thus comes from lightlike $x$. The main new simplification that arises in this case is
that for lightlike $x$ the OPE becomes an expansion 
in twist. Focusing on a generic CFT, the lowest twist operators are the identity and stress-tensor.\footnote{As discussed in the Introduction, light scalars with dimension $1 \leq \D \leq 2$ have twist lower than 2. They can thus potentially contribute to the commutator in the lightsheet limit. In appendix \ref{app:lssusy} we show in theories with $\cN=1$ supercoformal symmetry and no flavor symmetry that R-symmetry neutral light scalars are absent. Therefore only the identity and the stress tensor appear in the light sheet limit of the stress tensor commutator.} 
Thus we get schematically
\begin{align}
\lim_{\vv \to 0}[\cO_i(u,\vv,x_\perp),\cO_j(0)] =\lim_{\vv \to 0} \<[\cO_i(u,\vv,x_\perp),\cO_j(0)]\> + \sum_i f_{ij }^{\a\b}(u,x_\perp,\p)T_{\a\b}(0)+O(\vv)\, .
\end{align}
We will use this in section \ref{sec:LSC} to derive light sheet commutators.

\subsection{Equal-time stress-tensor commutators in QFT \label{sec:22}}

We now briefly summarize constraints on equal-time commutators of components of the stress-tensor which follow from the
usual Poincar\'e algebra 
\begin{align*}
[J_{\m\n},J_{\a\b}] &= i\left( J_{\m \b}\eta_{\n \a} + J_{\n \a}\eta_{\m \b} - J_{\n \b}\eta_{\m \a} - J_{\m \a}\eta_{\n \b}\right)\, ,\\
[P_\m,J_{\n \a}] &= i\left( \eta_{\m \n}P_\a - \eta_{\m \a}P_\n \right)\, ,\\
[P_\m,P_\n] &= 0.
\end{align*}
As is well-known, generators of the Poincar\'e algebra can be written as suitable integrals of the stress-tensor over equal-time surfaces \cite{SchwingerNonAbelian}.
Therefore, the equal-time commutators of components of the stress-tensor must have a structure which reproduces the Poincar\'e algebra
after suitable integration. This is by no means sufficient to completely determine the equal-time commutators. In particular, we obtain no
information about total derivative contributions to equal-time commutators, which could even be UV-divergent (these divergences are not
problematic for the Poincar\'e algebra, as we are instructed to first do all the relevant integrals before taking $\epsilon$ to zero). 

For example, the equal-time commutator of $T^{00}$ with itself must have the form\footnote{When discussing equal-time stress-tensor commutators we occasionally drop the time argument and denote $T^{\m\n}(\vec{x})=T^{\m\n}(0,\vec{x})$ to avoid clutter.} \cite{SchwingerNonAbelian}
\begin{align}
i[T^{00}(\vec{x}), T^{00}(\vec{y})] &= \left(T^{0k}(\vec{x})+T^{0k}(\vec{y})\right)\p_k \d(\vec{x}-\vec{y}) + \ldots
\end{align}
where the dots represent possibly UV-divergent total derivative terms.
Upon integrating over $\vec{y}$ we obtain
\begin{align}
i[T^{00}(\vec{x}), P^0] = \p_k T^{0k}(\vec{x})\, ,
\end{align}
which is simply the statement of conservation of energy
\begin{align}
\p_k T^{0k}(\vec{x}) + \p_0 T^{00}(\vec{x})=0\, .
\end{align}
The full Poincar\'e algebra determines the following equal-time commutators \cite{BoulwareDeser}
\begin{align*}
i[T^{00}(\vec{x}), T^{00}(\vec{y})] &= \left(T^{0k}(\vec{x})+T^{0k}(\vec{y})\right)\p_k \d(\vec{x}-\vec{y}) +\ldots\\
i[T^{00}(\vec{x}), T^{0m}(\vec{y})] &= \left(T^{mn}(\vec{x})+T^{00}(\vec{y})\d^{mn}\right)\p_n \d(\vec{x}-\vec{y}) +\ldots\\
i[T^{00}(\vec{x}), T^{mn}(\vec{y})] &= \left(-\p^0T^{mn}(\vec{x})+T^{0m}(\vec{y})\p^{n}+T^{0n}(\vec{y})\p^{m}\right) \d(\vec{x}-\vec{y}) +\ldots\\
i[T^{0k}(\vec{x}), T^{0m}(\vec{y})] &= \left(T^{0m}(\vec{x})\p^k+T^{0k}(\vec{y})\p^m\right) \d(\vec{x}-\vec{y}) +\ldots\\
i[T^{0k}(\vec{x}), T^{mn}(\vec{y})] &= \left(T^{mn}(\vec{x})\d^{kl}-T^{ml}(\vec{y})\d^{nk}-T^{nl}(\vec{y})\d^{mk}\right)\p_l \d(\vec{x}-\vec{y}) +\ldots
\end{align*}
where the ellipsis denote so-called Schwinger terms $\t^{\m\n,\a\b}(\vec{x},\vec{y})$ which, as discussed above, need to be total derivatives. In general, they can be UV divergent and are theory dependent. 

The commutator $[T^{kl}(\vec{x}), T^{mn}(\vec{y})]$ with all spatial indices is not fixed by
Poincar\'e symmetry as these components are not involved in any conservation equation. We will study these commutators using our first
principles approach in section \ref{STcomm4d}.

\section{Free scalar}
\label{sec:fs}

In this section we illustrate some of the ideas with the example of a free scalar field in four dimensions. The Lagrangian is given by 
\be \cL= -{1\over 2}\p^\m \phi \p_\m \phi \, ,
\ee
while the field can be expanded as usual 
\begin{align}
\label{fim}
\phi(x) = \int {d^3 p \over (2\pi^3)} {1\over \sqrt{2\om_p}} \left( a_{\vec{p}} e^{-i \om_p t + i\vec{p}\cdot \vec{x} } + a_{\vec{p}}^\dagger e^{i \om_p t -i \vec{p}\cdot \vec{x} } \right)\, .
\end{align}
where $\om_p = |\vec{p}|.$ The commutator is computed using $[a_{\vec{p}},a_{\vec{q}}^\dagger] = (2\pi)^3 \d (\vec{p}-\vec{q})$, and we obtain 
\begin{align}
[\phi(t,\vec{x}),\phi(0)]
={i \over 4\pi r}    \left( \d \left(t + r\right)   - \d \left(t - r\right) \right)\, ,
\end{align}
At equal times, the commutator of the time derivative of the field with the field itself is given by  
\begin{align}
\label{pdpe}
[\dot{\phi}(0,\vec{x}),\phi(0)] =-i \d(\vec{x})\, .
\end{align}
We reproduce these results using our prescription.

\subsection{Free scalar field commutator}
We derive these results using the OPE to illustrate our framework. The starting point is the OPE of two free scalars, which is given by
\begin{align}
\phi(x)\phi(0) = C_0x^{-2} + \sum_{n,l}C_{n,l} x^{2n +l} \left([\phi\phi]_{n,l}(0)+\text{descendants}\right)\, .
\end{align}
The ``double twist" operators $[\phi\phi]_{n,l}$ ($n,l$ are non-negative integers) do not contribute to the commutator since they have no branch cuts or singularities. 
As we already discussed, the identity contribution can be extracted from the two-point function, which reads explicitly
\begin{align}
\<\phi(t, \vec{x}) \phi(0, 0)\> = {1\over 4\pi^2}(-t^2 + r^2)^{-1}\, ,
\end{align}
where the normalization is fixed by (\ref{fim}). To order the operators in the Wightman functions, we use the usual $i\epsilon$ prescription \cite{Hartman:2015lfa}, such that the two different orderings are given by
\begin{equation}\label{eq:usefullscalar}
\begin{aligned}
\<\phi(t, \vec{x}) \phi(0) \> &\equiv \lim_{\eps \to 0}g^-_\eps (t,\vec{x})= \lim_{\eps \to 0} {1\over 4\pi^2} (-(t-i\eps)^2 + r^2)^{-1}\, ,\\
\<\phi(0) \phi(t, \vec{x})\> &\equiv \lim_{\eps \to 0} g^+_\eps (t,\vec{x}) = \lim_{\eps \to 0} {1\over 4\pi^2} (-(t+i\eps)^2 + r^2)^{-1}\, .
\end{aligned}
\end{equation}
As discussed in the previous section, we define the commutator $g^c(t,\vec{x})$ as a distribution in $(t,\vec{x})$. It is obtained by integrating against an arbitrary test function $f(t)$ as 
\begin{align}
\label{gce}
\int_{-\infty}^\infty dt \, g^c (t,\vec{x})f(t)&\equiv \lim_{\eps \to 0} \int_{-\infty}^\infty dt \left(g^-_{\eps}(t,\vec{x}) - g^+_{\eps} (t,\vec{x})\right)f(t)\, .
\end{align}
The computation now follows the general discussion and we can simply use the general result in (\ref{jan1}) with $h=1$. This leads to
\begin{align}
\label{fbcfe}
g^c (t,\vec{x}) = {i \over 4 \pi r} \left( \d \left(t+r\right)   - \d \left(t-r\right) \right),
\end{align}
which is indeed the expected commutator.

\subsection{Equal-time commutator}

To obtain the familiar canonical commutation relations for the free scalar, we start from (\ref{gce}) and set $t=0$. This yields 
\begin{align}
[\phi(0, \vec{x}), \phi(0,0)] = \lim_{\eps \to 0} \left( g_\eps^-(0,\vec{x})- g_\eps^+(0,\vec{x})\right) =0\, .
\end{align}
For the momentum-scalar commutator we then get
\begin{align}
\lim_{\eps \to 0} i\left(\p_t g_\eps^-(0,\vec{x})-\p_t g_\eps^+(0,\vec{x})\right) = \lim_{\eps \to 0}{\eps\over \pi^2(\eps^2+x_1^2+x_2^2+x_3^2)^2}\, .
\end{align}
This is a special case of (\ref{jan3}) with $h=2$, $p=1$, and $d=4$, so that $p=2h+1-d$ and the answer will be proportional to a delta function.
To verify this explicitly in this simple example, we integrate against a test function $f(\vec{x})$ where 
the only contribution comes from the origin in the $\eps \to 0$ limit
\begin{align}
\int d^3 x{\eps f(x_1,x_2,x_3)\over \pi^2(\eps^2+x_1^2+x_2^2+x_3^2)^2} &\simeq f(0,0,0) \int_{-\infty}^\infty dx_3\int_{-\infty}^\infty dx_2 \int_{-\infty}^\infty dx_1{\eps \over \pi^2(\eps^2+x_1^2+x_2^2+x_3^2)^2}\, ,\nonumber\\
&= f(0,0,0)\, .
\end{align}
We have now reproduced the usual canonical commutation relation (\ref{pdpe}).

\section{Stress-tensor commutators in CFT$_2$}
\label{sec:stcft2}

Two-dimensional CFTs are very special due to the infinite-dimensional Virasoro symmetry. This implies that we have complete control over the OPE of two stress-tensors, and more generally of the OPE
of a stress-tensor with any other primary operator. The OPE of two holomorphic stress-tensors is universal and all terms are fixed by a unique number, the central charge $c$. The regular terms in the OPE have positive integer power, and they do not contribute to the commutator when continued to Lorentzian signature. The usual Euclidean OPE is given in terms of coordinates  $z=x_1 - i \t,~\zb=x_1 + i \t$ where $(\t,x_1)$ parametrize the Euclidean plane. Wick rotating to Lorentzian time as $\t \to it$ and defining lightcone coordinates $u=t-x_1,~\vv=t+x_1~$
the OPE reads
\begin{align}
T_{\vv\vv}(\vv)T_{\vv\vv}(\vv') =  {c\over 8\pi^2(\vv-\vv')^4} - {T_{\vv\vv}(\vv')\over \pi(\vv-\vv')^2} -{\p_{\vv'} T_{\vv\vv}(\vv') \over 2\pi(\vv-\vv')}+\dots \, .
\end{align}
We now apply our general methods to compute the commutator from the OPE. We show explicitly the computation for the identity contribution here while the rest of the computation is shown in appendix \ref{app:vir2}.

We set $\vv'=0$ and define the identity contribution to the commutator $g_0^c(\vv)$ as a distribution in $\vv$ by using the appropriate $i\epsilon$ prescription. Integrating against a test function $f(\vv),$ the commutator $g^c_0 (\vv)$ is obtained as
\begin{align}
\label{ct2dmt}
\int_{-\infty}^\infty d\vv~ g^c_0 (\vv)f(\vv) &\equiv \lim_{\eps \to 0} \int_{-\infty}^\infty d\vv~ \left({c\over 8\pi^2(\vv- i\eps)^4}
- {c\over 8\pi^2(\vv+ i\eps)^4}
\right)f(\vv)\, .
\end{align}
We perform the integral over the real $\vv$ line on the right hand side of (\ref{ct2dmt}) by deforming the contour. The first term has a pole on the upper half plane and the second term has a pole on the lower half plane. The outcome of using the two different contours agree in the $\eps \to 0$ limit and give 
\begin{align}
\label{g0c}
g^c_0 (\vv) =-{ic\over 24 \pi}\p^3_\vv \d(\vv)\, .
\end{align}
Performing a similar computation for the rest of the terms (cf appendix \ref{app:vir2} for details), we get
\begin{align}
\label{st2d}
[T_{\vv\vv}(\vv), T_{\vv\vv}(\vv')] = i\left(T_{\vv\vv}(\vv) + T_{\vv\vv}(\vv')\right) \p_\vv \d(\vv-\vv')  -{ic\over 24 \pi}\p^3_\vv \d(\vv-\vv')\, ,
\end{align}
which is the standard commutator form of the Virasoro algebra.

\subsection{Virasoro generators as light-ray operators}

The usual construction of the generators of the Virasoro algebra from the OPE follows by taking $L_n \sim \oint z^{n+1} T(z)$, which have
a natural interpretation through radial quantization in terms of a CFT on an Euclidean cylinder. It is therefore not entirely obvious
how one should describe the Virasoro generators on the Lorentzian plane. One might be tempted to consider generators in  the form
of light-ray operators $L_n \sim \int du \, u^{n+1} T_{uu}$ but these operators do not in general have well-defined matrix 
elements\footnote{By looking at three-point functions of $T_{\mu\nu}$ with two other primaries we find that matrix elements for large $u$ 
	in $d$ dimensions typically behave as $T_{uu}\sim u^{-d-2}$, $T_{u\vv}\sim u^{-d}$, $T_{\vv\vv}\sim u^{-d+2}$, $T_{uA}\sim u^{-d-1}$, $T_{\vv  A}\sim 
	u^{-d+1}$ and $T_{AB}\sim u^{-d}$.}. We can however pull-back the usual construction of Virasoro generators from the cylinder to the plane. 
This does not rely on radial quantization, but on the observation that we can map the plane to a diamond shaped region on the Lorentzian
cylinder. This proceeds through a conformal transformation which maps the metric of the plane $\eta$ to 
$g=\Omega^2 \eta$ where the conformal factor reads
\begin{align}
\Omega^2 = 4(1+\vv^2)^{-1} (1+u^2)^{-1}\, .
\end{align}
This embedding will in particular allow us to extend the domain of our operators of interest beyond the spacetime infinities of the original plane. 
The detailed embedding map is given by
\begin{align}
T &= \tan^{-1}\vv + \tan^{-1}u,\\
\Phi &= \tan^{-1}\vv - \tan^{-1}u.
\end{align}
where the coordinates $T,\Phi$ are restricted to the range
\begin{align}
-\pi < T+\Phi < \pi\, , \qquad \qquad -\pi < T-\Phi < \pi\, .
\end{align}
\begin{figure}[t!]
	\centering
	\adjincludegraphics[trim={7cm 9.7cm 7cm 9.7cm},clip]{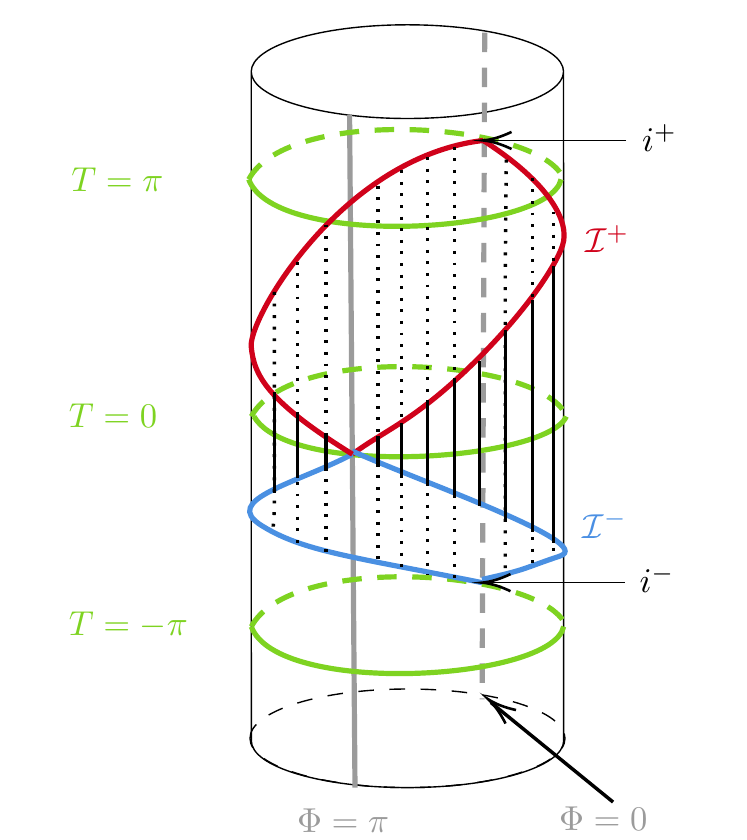}
	\caption{Embedding of the Minkowski plane on the Lorentzian cylinder.}
	\label{figFeynman}
\end{figure}
In these coordinates, the metric $g$ reads
\begin{align}
ds^2_g = -dT^2 + d\Phi^2 \, . 
\end{align}
We identify $\Phi=\pi = -\pi$ which corresponds to spatial infinity of the Lorentzian plane; the metric $g$ describes the Lorentzian cylinder, see Figure \ref{figFeynman}. Next we define coordinates
\begin{align}
U={T-\Phi}=2 \tan^{-1}u,~~~V={T+\Phi}=2 \tan^{-1}\vv\, .
\end{align}
We can now compute the stress-tensor in this new set of coordinates. The stress-tensor transforms as\footnote{Curly brackets denote the Schwarzian derivative $\{f;g\}={(d^3f/dg^3)\over (df/dg)} -{3\over 2}\left({d^2f/dg^2\over df/dg}\right)^2$}
\begin{align}
T_{VV}(V) = \left( {dV\over d\vv} \right)^{-2} T_{\vv\vv}(\vv) + {c\over 12} \{\vv;V\}\, .
\end{align}
The Virasoro generators are defined as usual, and we get 
\begin{align}
L_n = \int_{-\pi}^{\pi}dV e^{inV}T_{VV}(V) &= { c\pi \over 12}\d_{n,0}+{1\over 2} \int_{-\infty}^{\infty}d\vv (1+\vv^2) \exp \left[2in \tan^{-1}\vv\right] T_{\vv\vv}(\vv)\, ,\\
&= { c\pi \over 12}\d_{n,0}-{1\over 2} \int_{-\infty}^{\infty}d\vv (i+\vv)^{1-n} (i-\vv)^{1+n} T_{\vv\vv}(\vv)\, .\label{ln2dnd}
\end{align}
In this definition the coordinates are taken to be dimensionless. It is convenient to include an explicit length scale $R \in \mathbb{R}_{>0}$ which is the size of the circle, in the generators 
such that they are dimensionless. Neglecting the vacuum energy, we define our Virasoro generators as 
\begin{align}
\label{ln2d}
L_n \equiv -{1\over 2R} \int_{-\infty}^{\infty}d\vv (iR+\vv)^{1-n} (iR-\vv)^{1+n} T_{\vv\vv}(\vv)\, .
\end{align}
Note that $L_n^\dagger = L_{-n}.$
Integrating the stress-tensor commutator (\ref{st2d}), it is seen the generators defined in \eqref{ln2d} indeed obey the standard Virasoro algebra,
\begin{align}
[L_m, L_n] =(m-n)L_{m+n} + {(m^3-m)c\over 12}\d_{m+n,0}\, . \label{Virasoro2}
\end{align}
We refer the reader to appendix \ref{app:vir2} for details.

We continue with the correlators of (\ref{ln2d}). The three-point function of the stress-tensor with two primaries is given as 
\begin{align}
\<T_{\vv\vv}(\vv)\mathcal{O}(\vv_1) \mathcal{O}(\vv_2)\> = {h\over (\vv_1-\vv_2)^{2h-2} (\vv-\vv_1)^{2} (\vv-\vv_2)^{2}}\, ,\label{localT3pt}
\end{align}
where $h$ is the conformal dimension of the scalar operator $\mathcal{O}$. To order the operators, we need to enforce the appropriate $i\epsilon$ prescription. To order them as seen above, we give $\vv_1$ and $\vv_2$ small positive imaginary parts with $\Im(\vv_1)< \Im(\vv_2)$. To obtain $T_{\vv\vv}$ acting on the right vacuum $\<O(\vv_1) O(\vv_2)T_{\vv\vv}(\vv)\>$ we give $\vv_1$ and $\vv_2$ small negative imaginary parts with $\Im(\vv_1) <\Im(\vv_2)$.  The configuration $\<O(\vv_1) T_{\vv\vv}(\vv)  O(\vv_2)\>$ is obtained by giving $\vv_1$ a negative and $\vv_2$ a positive imaginary part. To obtain the three-point function involving light-ray operators, we need to perform the following integral 
\begin{align}
\braket{L_n\mathcal{O}(\vv_1)\mathcal{O}(\vv_2)} = -{1\over (\vv_1-\vv_2)^{2h-2}}\int_{-\infty}^\infty d\vv {h (iR-\vv)^{n+1}\over 2R (iR+\vv)^{n-1} (\vv-\vv_1)^{2} (\vv-\vv_2)^{2}}  \, ,
\end{align}
We perform this integral with residues. The integrand dies off along the positive and negative imaginary axes so we can close the contour on either side. 

The subset $n=-1,0,1$ is special; the only singularities of the integrand are located at $\vv=\vv_1,\vv_2.$ We obtain, for the ordered correlators
\be 
\label{2dlng}
\braket{L_n\mathcal{O}(\vv_1)\mathcal{O}(\vv_2)} = \braket{\mathcal{O}(\vv_1)\mathcal{O}(\vv_2)L_n} = 0\, , \qquad n = -1,\, 0,\, 1\, .
\ee
For $n\geq 2$ the integrand has an additional pole at $\vv=-iR$ and for $n\leq -2$ the integrand has an additional pole at $\vv=iR.$ Thus

\begin{align}
\braket{\mathcal{O}(\vv_1)\mathcal{O}(\vv_2)L_n}& =0 \, \qquad \text{for } n\geq 2\, ,\\
\braket{L_n\mathcal{O}(\vv_1)\mathcal{O}(\vv_2)} &=0\, \qquad \text{for } n\leq -2\, .
\end{align}
These results combined with (\ref{2dlng}) imply
\begin{align}
\label{ln2dv}
L_n|0\>=0,\qquad ~~n\geq -1\, ,
\end{align}
as usual for Virasoro generators.

We now compute $\< \mathcal{O}(\vv_1) L_{n} \mathcal{O}(\vv_2)\>.$ For $n\geq 2$, the integrand has poles in the lower half plane at $\vv=-iR,\vv_1$ and in the upper half plane at $\vv=\vv_2$, while for $n = -1,\, 0,\, 1$, the integrand has poles in the lower half plane at $\vv=\vv_1$ and in the upper half plane at $\vv=\vv_2.$  Closing the contour in the upper half plane we get
\begin{align}
\braket{\mathcal{O}(\vv_1)L_n \mathcal{O}(\vv_2)}=\frac{2 i \pi h (R+i\vv_2)^n\left(R^2+i Rn(\vv_1-\vv_2)+\vv_1\vv_2\right)}{R(R-i \vv_2)^n(\vv_1-\vv_2)^{2h+1}},~~n \geq -1\, .
\end{align}
Using
\begin{align}
\braket{\mathcal{O}(\vv_1) L_n \mathcal{O}(\vv_2)} = \<\mathcal{O}(\vv_1) [L_n, \mathcal{O}(\vv_2)]\>, ~~\quad n\geq -1\, ,
\end{align}
this implies
\begin{align}
[L_n, \mathcal{O}(\vv)] = i \pi  {(R+i\vv)^n\over R(R-i \vv)^n } \left( 2h (iRn+\vv) + (R^2+\vv^2)\p_{\vv}\right)\mathcal{O}(\vv)\ .\label{greg1}
\end{align} 
Actually, due to the fact that $L_{-n} = L_n^\dagger$, \eqref{greg1} holds for any $n$.

\subsubsection{$SL(2,\mathbb{R})$ subalgebra}

We now identify the usual $SL(2,\mathbb{R})$ subalgebra of Virasoro symmetry. Let us denote
\begin{align} \label{jan5}
\tilde{L}_k = \int_{-\infty}^\infty d\vv\, \vv^{k+1} T_{\vv\vv}(\vv),\qquad \qquad k=-1,0,1\, .
\end{align}
The three-point functions involving the operators \eqref{jan5} can be computed by direct integration, and we find, for example 
\begin{align}
\< \cO(\vv_1) \tilde{L}_{-1} \cO(\vv_2)\> &= 4\pi ih{1\over (\vv_1-\vv_2)^{2h+1}}\, .
\end{align}
Using
\begin{align}
\< \cO(\vv_1) \tilde{L}_k \cO(\vv_2)\> = {1\over 2} \< \cO(\vv_1) [\tilde{L}_k, \cO(\vv_2)]|0\> - {1\over 2} \< [\tilde{L}_k,\cO(\vv_1)] \cO(\vv_2)\>\, ,
\end{align}
we derive the action of \eqref{jan5} on primary scalar operators as 
\begin{align}
[\tilde{L}_{-1}, \cO(\vv)] &= 2\pi i \p_{\vv}\cO(\vv)\, ,\\
[\tilde{L}_0, \cO(\vv)] &= 2\pi i  \left(h+\vv\p_{\vv}\right)\cO(\vv)\, ,\\
[\tilde{L}_1, \cO(\vv)] &= 2\pi i \left(2 h \vv+\vv^2 \p_{\vv}\right)\cO(\vv)\, .
\label{jan4}
\end{align}
In addition, the operators defined in \eqref{jan5} are mapped to our $L_n$ \eqref{ln2d} as
\begin{align}
\tilde{L}_{-1} &= {1\over 2R}\left( 2L_0+L_1+L_{-1}\right)\, ,\\
\tilde{L}_0 &= {i\over 2}\left( L_{-1}-L_{1}\right)\, ,\\
\tilde{L}_1 &= {R\over 2}\left( 2L_0-L_1-L_{-1}\right)\, .
\end{align}
This shows that the familiar $SL(2,\mathbb{R})$ subalgebra of the Virasoro algebra is generated by our $L_{-1},\, L_{0},\, L_1$ that are defined in \eqref{ln2d}.

Incidentally the two-dimensional ANEC operator is given by
\begin{align}
\cE=\int_{-\infty}^{\infty}d\vv T_{\vv\vv}(u,\vv)={1\over 2R}\left(2 L_0 + L_1 +  L_{-1}\right)\, ,
\end{align}
which is independent of $u.$ The commutativity of ANEC operators follows trivially from the Virasoro algebra \eqref{Virasoro2} and we conclude \cite{Kologlu:2019bco}
\be [\cE,\cE]=0\, .
\ee
In addition, the positivity of the ANEC operator is the statement that \cite{Faulkner:2016mzt,Hartman:2016lgu,de_Boer_2020}
\be \langle \psi | 2L_0 + L_1 + L_{-1} | \psi\rangle \geq 0\, ,
\ee for any state $\psi$ in a two-dimensional CFT.

In summary, we have seen that the light-ray operators $\tilde{L}_k$ in (\ref{jan5}) act
on primaries in a way which is very similar to the action of the corresponding generators on the Euclidean plane in radial
quantization. However, in contrast to the Euclidean case, the $\tilde{L}_k$ are not well-defined operators for $k\neq -1,0,1$.
There exist a different set of generators $L_n$ for all $n$, that we introduce in equation \eqref{ln2d}, which do form a bona fide Virasoro algebra. Their three-point functions are well-defined for any $n\in \mathbb{Z}$ and their $SL(2,\mathbb{R})$ subalgebra agrees with the algebra generated by $\tilde{L}_k$ for $k=-1,0,1$.

\section{Stress-tensor commutators in CFT$_4$\label{STcomm4d}}
In this section, we investigate commutators of various components of the stress-tensor in four-dimensional CFTs. The main difference from two dimensions is the fact that the stress-tensor OPE in CFT$_4$ is not universal, and in practice, any (neutral) operator can appear. We will analyse two simplifying limits: the equal-time and light sheet limits. In both cases, we derive explicit expressions for the commutators of interest. The identity contribution is computed from the stress-tensor two-point function $\<[T,T]\>$, and for the stress-tensor contribution we analyse the three-point function $\<[T,T]T\>$. As discussed in the Introduction the latter encodes the entire conformal family of the stress-tensor, albeit it is in practice difficult to extract this information. Our method is to compute various integrals of $\<[T,T]T\>$ and thereby compute $[T,T]\sim T$ up to total derivatives.

To compute the commutator $[T_{\mu \nu}(x), T_{\sigma \rho}(y)]$ for generic causally related points $x$ and $y$, we can use the same method as in two dimensions, and use (\ref{cd}). The relevant information is encoded in the OPE \cite{Osborn:1993cr}
\begin{align}
\label{ttope4d}
T_{\mu \nu}(x) T_{\sigma \rho}(0) \sim C_{T} \frac{\mathcal{I}_{\mu \nu, \sigma \rho}(x)}{x^{2 d}}&+{t_{\mu \nu \sigma \rho \alpha \beta}(x)\over x^d} T_{\alpha \beta}(0)+B_{\mu \nu \sigma \rho \alpha \beta \lambda}(x) \partial_{\lambda} T_{\alpha \beta}(0)+\ldots \no\\
&+\sum_{i} {}^{i}C_{\m\n\s\r}^{\a\b\ldots}(x,\p)O_{\a\b\ldots}^{i}(0)\, ,
\end{align}
In the first line of \eqref{ttope4d} we display the identity and stress-tensor contributions while the second line of \eqref{ttope4d} captures the contribution from other primary operators.

The identity contribution is the simplest and can be extracted directly from the stress-tensor two-point function, which is completely fixed by conformal symmetry up to a number $C_T$, and is given as \cite{Osborn:1993cr}
\begin{align}
\label{tt24d}
\langle T^{\mu \nu}(x) T^{\sigma \rho}(0)\rangle&=\frac{C_{T}}{x^{2 d}} \mathcal{I}^{\mu \nu, \sigma \rho}(x)\, ,
\end{align}
with
\begin{align} 
I^{\m\n}(x)= \d^{\m \n} -2 {x^\m x^\n \over x^2}\, ,\qquad 
\mathcal{I}^{\mu \nu, \sigma \rho}(x)=\frac{1}{2}\left(I^{\mu \sigma}(x) I^{\nu \rho}(x)+I^{\mu \rho}(x) I^{\nu \sigma}(x)\right)-\frac{1}{d} \delta^{\mu \nu} \delta^{\sigma \rho}\, .
\end{align}
To carry out our computations, we define the following tensor
\begin{align}
\label{gx}
\TTten^{\m\n\s\r}(x) &\equiv {\< T^{\mu \nu}(x) T^{\sigma \rho}(0)\>\over C_T}\, .
\end{align}
The commutator is then obtained by taking the $\epsilon \rightarrow 0$ limit of 
\begin{align}
\TTten_\eps^{\m\n\s\r}(x) &\equiv \TTten^{\m\n\s\r}(t-i\eps, \vec{x}) - \TTten^{\m\n\s\r}(t+i\eps, \vec{x})\, ,\label{eq:ge4d}
\end{align} 
as 
\begin{align}
\TTten^{[\m\n,\s\r]}(x) &\equiv \lim_{\eps \to 0} \TTten_\eps^{\m\n\s\r}(x)\, .
\end{align}
The two sides of this equation are understood to be equal as distributions in $x.$ In equation \eqref{eq:ge4d}, the Lorentzian separation $i\epsilon$ enforces the correct ordering of the operators such that we obtain a commutator. This is in close analogy to \eqref{eq:usefullscalar} in the free scalar case.

We explain the details in an example. For the case where $\mu=\nu=\sigma=0$ and $\rho=i$, the two-point function is given by
\begin{align}
\label{t000ie}
\TTten^{000i}(x) = {2t x_i \over x^{10}} + {4t^3 x_i \over x^{12}}\, . 
\end{align}
We extract the commutator from an integral against a test function $f(x)$ 
\begin{align}
\int d^4x~ \TTten^{[\m\n,\s\r]}(x) f(x) &\equiv \lim_{\eps \to 0} \int d^4x~ \TTten_\eps^{\m\n\s\r}(x) f(x)\, .
\end{align}
This yields
\begin{equation}
\begin{aligned}
\TTten^{[00,0i]}(x)= &{i\pi x_i \over 32 r^7} \left( \d'(t+r)-\d'(t-r) \right)  - {i\pi x_i \over 32 r^6} \left( \d''(t+r)+\d''(t-r) \right) + {i\pi x_i \over 64 r^5} \left( \d^{(3)}(t+r)-\d^{(3)}(t-r) \right) \\
& - {i\pi x_i \over 192 r^4} \left( \d^{(4)}(t+r)+\d^{(4)}(t-r) \right) + {i\pi x_i \over 960 r^3} \left( \d^{(5)}(t+r)-\d^{(5)}(t-r) \right)\, .
\end{aligned}
\end{equation}
For generic $x$ we cannot say much more. We will then focus on two physically interesting and simpler kinematics: the equal-time limit and the light sheet limit. In both cases we will analyse the identity and stress-tensor contributions separately, and find that the identity contribution is divergent in both cases.

\subsection{Equal-time commutators}

As discussed in section \ref{sec:22}, equal-time commutators of stress-tensors are partially fixed by Poincar\'e symmetry. For commutators $[T^{\m\n}(x^0,\vec{x}),T^{\a\b}(x^0,0)]$ involving at least one time component, conservation laws fix the stress-tensor contribution up to total derivatives. The identity contribution is not fixed, and is expected to be divergent on dimensional grounds and by the spectral decomposition in QFT \cite{BoulwareDeser}. Nevertheless, causality tells us generic commutators are supported on and inside the lightcone, which shrinks to a point in the equal-time limit. We thus expect equal-time commutators to be given by spatial delta functions and derivatives thereof. Using our method discussed above we perform the explicit computation and verify this.

\subsubsection{Identity contribution}

The $t\rightarrow 0$ limit of \eqref{eq:ge4d} places the two stress-tensors on a time slice, and our starting point for the equal-time commutator is
\begin{align}
\TTten_\eps^{\m\n\s\r}(\vec{x}) &\equiv \TTten^{\m\n\s\r}(-i\eps, \vec{x}) - \TTten^{\m\n\s\r}(i\eps, \vec{x})\, ,
\end{align}
The updated argument of the tensor $\TTten_\eps^{\m\n\s\r}$ reflects the fact that we are computing an equal-time commutator, which is a distribution in the three-dimensional vector $\vec{x}.$  We obtain the commutator as 
\begin{align}
\TTten^{[\m\n,\s\r]}(\vec{x}) &= \lim_{\eps \to 0} \, \TTten_\eps^{\m\n\s\r}(\vec{x})\, .
\end{align}
In generic QFT, the spectral decomposition implies the identity contribution to stress-tensor commutators is non-zero only if an odd number of time components are involved \cite{BoulwareDeser}. The same conclusion is reached in CFT observing that the two-point function (\ref{tt24d}) is odd in time for the same class of commutators. This implies that only $\TTten^{[00,0i]}(\vec{x})$ and $\TTten^{[0i,jk]}(\vec{x})$ are non-zero. We compute these in turn.

\subsection*{$\boldsymbol{\<[T^{00}(0,\vec{x}),T^{0i}(0)]\>}$}

Starting from \eqref{t000ie} and using our prescription, the commutator is given by the distributional equation
\begin{align}
\TTten^{[00,0i]}(\vec{x}) =\lim_{\eps \to 0} \left[-{4i x_i \eps \over (\eps^2+x_1^2+x_2^2+x_3^2)^5} + {8i x_i \eps^3 \over (\eps^2+x_1^2+x_2^2+x_3^2)^6}\right]\, .
\end{align}
By dimensional analysis this can only be a combination of $\lim_{\eps \to 0} {1\over \eps^2} \p_i \D \d(\vec{x})$ and $\p_i \D^2 \d(\vec{x}).$ Integrating against suitable test functions we determine
\begin{align}
\label{t000i}
\TTten^{[00,0i]}(\vec{x}) = {i\pi^2\over 480} \p_i \D^2 \d(\vec{x})+\lim_{\eps \to 0} \left[ {i\pi^2\over 240\eps^2} \p_i \D \d(\vec{x})  \right]  \, .
\end{align}
where $\D= \partial_i \partial^i$ denotes the three-dimensional Laplacian.

\subsection*{$\boldsymbol{\<[T^{0i}(0,\vec{x}),T^{jk}(0)]\>}$}

We extract from the two-point function
\begin{align}
\TTten^{[0i,jk]}(\vec{x})  =\lim_{\eps \to 0} \left[-{8i x_i x_j x_k \eps \over (\eps^2+x_1^2+x_2^2+x_3^2)^6}\right]\, .
\end{align}
This is a combination of $\lim_{\eps \to 0} {1\over \eps^2} \p_i \p_j\p_k  \d(\vec{x})$ and $\p_i \p_j\p_k \D \d(\vec{x}).$ Integrating against suitable test functions we determine
\begin{align}
\label{t0ijk}
\TTten^{[0i,jk]}(\vec{x}) = {i\pi^2\over 480} \p_i \p_j \p_k \D \d(\vec{x})+\lim_{\eps \to 0} \left[ {i\pi^2\over 240\eps^2} \p_i \p_j \p_k \d(\vec{x})  \right]\, .  
\end{align}
The results (\ref{t000i}) and (\ref{t0ijk}) are to be contrasted with equations (25) and (26) in \cite{Mahanthappa:1969bg}, where a distributional analysis \cite{Brandt} of vacuum expectation values of equal-time stress-tensor commutators in a generic QFT is performed using the spectral decomposition. The structure of the finite and UV divergent pieces match; this is expected due to locality, nevertheless highlights the role of the Euclidean time displacement $\eps$ as a UV regulator in equal-time commutators.

\subsubsection{Stress-tensor contribution}
\label{sec:etstc}

The stress-tensor contribution to the commutator is more complicated. The strategy that we will use is the following: Compute the two-point function of the commutator of two stress-tensors with a third stress-tensor as 
\begin{align}
\label{tttxd}
\< [T^{\m \n}(0,\vec{x}), T^{\s \r}(0)] T^{\a\b}(w)\>\, .
\end{align}
where $w$ is an arbitrary spacetime point. We obtain this expression from the stress-tensor three-point function. Viewed as a distribution in $\vec{x},$ (\ref{tttxd}) is best understood by integrating it against a test function $f(\vec{x})$
\begin{align}
\label{etimp}
\int d^3x f(\vec{x})\< [T^{\m \n}(0,\vec{x}), T^{\s \r}(0)] T^{\a\b}(w)\>\, .
\end{align}
In principle, from this integral we can extract the stress-tensor contribution to $[T^{\m \n}(0,\vec{x}), T^{\s \r}(0)]$ as a distribution in $\vec{x}$ by matching this with its two-point function with $T^{\a\b}(w).$

In practice, the integral (\ref{etimp}) for arbitrary $f(\vec{x})$ is too complicated to perform. Therefore we resort to choosing the simplest test function $f(\vec{x})=1.$\footnote{This is of course not a proper test function. However, the fact that we are actually computing the two-point function of the distribution in question with $T^{\a\b}(w)$ makes the integral finite.} This gives a manageable integral at the expense of all the information about total derivative contributions to $[T^{\m \n}(0,\vec{x}), T^{\s \r}(0)]$. If desired these terms can be systematically computed setting $f(\vec{x})$ to polynomials in components of $\vec{x}.$

We continue with the computation. Our starting point is the Wightman function 
\begin{align}
\TTTten^{\m\n\s\r\a\b}(x;w)&\equiv\< T^{\m \n}(x) T^{\s \r}(0) T^{\a\b}(w)\>\, ,
\end{align}
and we denote the integrated commutator as 
\begin{align}
\label{ecii}
\TTTten^{[\m\n,\s\r]\a\b}_\text{int.}(w) &\equiv\int d^3x\,  \<[T^{\m\n}(0,\vec{x}), T^{\s\r}(0)] T^{\a\b}(w)\> \, ,
\end{align}
where the subscript ``int." is there to remind us this is the integrated commutator.

The stress-tensor three-point function is fixed by conformal symmetry (up to three numbers) and is given as
\begin{align}
\label{ttt4d}
\TTTten^{\m\n\s\r\a\b}(x;w)
&={1\over x^4 (x-w)^4 w^4} \cI^{\m \n}_{~~\m'\n'}(x) \cI^{\s\r }_{~~\s'\r'}(w) t^{\m' \n' \s' \r' \a \b}(X_{wx})\, .
\end{align}
We refer the reader to \cite{Osborn:1993cr} for the definitions of the tensors appearing in (\ref{ttt4d}). The (integrated) commutator is then obtained by taking the $\epsilon\rightarrow 0$ limit of 
\begin{align}
\label{ge6}
\TTTten^{\m\n\s\r\a\b}_{\eps}(\vec{x};w) & \equiv \TTTten^{\m\n\s\r\a\b}(- i\eps,\vec{x};w) -\TTTten^{\m\n\s\r\a\b}(i\eps,\vec{x};w)\, ,
\end{align}
as 
\begin{align}
(TTT)^{[\m\n,\s\r]\a\b}_\text{int.}(w) &\equiv \lim_{\eps\to 0} \int d^3x~ \TTTten^{\m\n\s\r\a\b}_{\eps}(\vec{x};w)\, .
\end{align}
To ensure $T^{\a\b}(w)$ acts on the right vacuum we set Im$(w^0)>\eps>0.$ This assignment is crucial as we perform the spatial integrals. 

To illustrate the details of the computation we focus on the example of $[T^{00}(0,\vec{x}), T^{00}(0)].$ As discussed in section \ref{sec:22}, we expect the result to be 
\begin{align}
\TTTten^{[00,00]00}_\text{int.}(w) = -i\p_k \<T^{0k}(0)T^{00}(w)\>\, . 
\end{align}
To simplify the tensor $\TTTten^{000000}_{\eps}(\vec{x};w)$ we set, without loss of generality, $w=(w^0,0,w_2,0)$ such that $x \cdot w = \pm i\eps w^0 +x_2 w_2$. In addition, to be able to manage the spatial integral, we expand $\TTTten^{000000}_{\eps}(\vec{x};w)$ in powers of $x^2=r^2+\eps^2$ as 
\begin{align}
\label{lamin}
\TTTten^{000000}_{\eps}(\vec{x};w)&= {\TTTten^{000000}_{\eps,5}(\vec{x};w)\over (r^2+\eps^2)^5} + {\TTTten^{000000}_{\eps,4}(\vec{x};w)\over (r^2+\eps^2)^4} + {\TTTten^{000000}_{\eps,3}(\vec{x};w)\over (r^2+\eps^2)^3}\nonumber\\
& + {\TTTten^{000000}_{\eps,2}(\vec{x};w)\over (r^2+\eps^2)^2} + {\TTTten^{000000}_{\eps,1}(\vec{x};w)\over (r^2+\eps^2)} +\mathcal{O}((x^{2})^0)\, .
\end{align}
In this expansion, all the terms that are of $\mathcal{O}((x^{2})^{n})$ with $n\geq 0$ are such that they vanish in both the $\vec{x} \to 0$ and $\eps \to 0$ limits. We conclude that they are equal to the zero distribution in the $\eps \to 0$ limit. The terms that are explicitly written in (\ref{lamin}) all vanish in the $\eps \to 0$ limit, but they give $\eps^{-p}$ for some $p>0$ in the $\vec{x} \to 0$ limit; they are non-zero as distributions in $\vec{x}$ and are the terms that we need to keep track of to be able to obtain the commutators of interest.

The $\cO((x^{2})^{-1})$ term requires special treatment, since its integral over space formally diverges. We expand $\TTTten^{000000}_{\eps,1}(\vec{x};w)$ in powers of  $x_2$ and $\eps;$ it starts at $O(\eps^1 (x_2)^0).$ The relevant integral is then
\begin{align}
\int d^3 x {\eps (x_2)^n\over (r^2+\eps^2)^m} = \frac{\pi((-1)^n+1)\G\left(\frac{n+1}{2}\right)\eps ^{-2 m+n+4}\G\left(m-\frac{n}{2}-\frac{3}{2}\right)}{2 \G (m)}\, ,
\end{align}
which converges for $m>1,~m>{3\over 2}+{n\over 2}.$ We analytically continue the integral in $m$ and define its $m\to 1$ limit as\footnote{The integral of interest can equivalently be obtained by differentiating (\ref{acm}) with respect to $\eps.$}
\begin{align}
\label{acm}
\int d^3 x {\eps (x_2)^n\over r^2+\eps^2} = -\frac{\pi ^2}{n+1} \left((-1)^n+1\right)
\eps^{n+2}~ \text{sec}
\left(\frac{\pi n}{2}\right)\, .
\end{align}
This expression is finite for integer $n \geq 0$ and vanishes in the $\eps \to 0$ limit. This implies 
\be \lim_{\eps \to 0}{\TTTten^{000000}_{\eps,1}(\vec{x};w)\over (r^2+\eps^2)}= 0\, ,
\ee 
as a distribution in $\vec{x}.$ The rest of the terms combine to give
\begin{align}
\TTTten^{[00,00]00}_\text{int.}(w) = 2i C_T {\left[3(w^0)^5 + 14(w^0)^3 (w_2)^2 + 7 w^0 (w_2)^4\right]\over (-(w^0)^2+(w_2)^2)^7}\, .
\end{align}
This matches with $-i\p_k \<T^{0k}(0,x_1,x_2,x_3)T^{00}(w^0,0,w_2,0)\>|_{x_1=x_2=x_3=0}$. As we discussed, this is a consequence of Poincar\'e symmetry\footnote{To further illustrate the method and gain confidence in it in appendix \ref{app:etete} we compute the stress tensor contribution to $[T^{0k}(\vec{x}),T^{0m}(0)]$ upto total derivatives.}, however the techniques we developed will prove useful in computing light sheet commutators, to which we now turn.

\subsection{Light sheet commutators\label{sec:LSC}}

We continue with the computation of stress-tensor commutators in the light sheet limit \\$\lim_{\vv \to 0}[T^{\m\n}(u,\vv,x_\perp),T^{\a\b}(0)].$ As discussed in section \ref{sec:gencom}, only the identity operator and the stress-tensor contribute to commutators of local operators on a common light sheet, which is composed of lightlike and spacelike separated points. We extract these from the stress-tensor two-point function $\<[T,T]\>$ and three-point function $\<[T,T]T\>$ as in the equal-time case. We switch to lightcone coordinates where the commutator is computed from
\begin{align}
\label{cdlc}
[\cO_i(u,\vv,x_\perp),\cO_j(0)] = \lim_{\eps \to 0} \big[\cO_i(u-i\eps,\vv-i\eps,x_\perp) \cO_j(0) - \cO_i(u+i\eps,\vv+i\eps,x_\perp) \cO_j(0)\big]\, .
\end{align}

\subsubsection{Identity contribution}
In this subsection, we compute the identity contribution to light sheet commutators. As in the previous section this is extracted from the stress-tensor two-point function, now in lightcone coordinates $(u,\vv,x_\perp)$. For example we have 
\begin{align}
{\< T^{\vv\vv}(u,\vv,x_\perp) T^{\vv\vv}(0)\> \over C_T} = {4 \vv^4  \over x^{12}}\, ,
\end{align}
where $x^2 = -u \vv  + x_\perp^2$. 

To compute the commutator, we integrate it against an arbitrary test function as  
\begin{align}
\int du\,  
{\braket{[T^{\vv\vv}(u,\vv,x_\perp),T^{\vv\vv}(0)]}\over C_T}f(u) \equiv \lim_{\eps \to 0} \int\, du\left( {4 (\vv  - i\eps)^4  \over (x_\perp^2-(u- i\eps)(\vv- i\eps))^6} - {4 (\vv  + i\eps)^4  \over (x_\perp^2-(u+ i\eps)(\vv+ i\eps))^6}\right)f(u)\, ,
\end{align}
which gives
\begin{align}
\label{tvvvvi}
{\<[T^{\vv\vv}(u,\vv,x_\perp),T^{\vv\vv}(0)]\>\over C_T} = -{i\pi\over 15\vv^2} \d^{(5)} \Big(u-{x_\perp^2 \over \vv}\Big)\, .
\end{align}
Note that the delta function appearing in \eqref{tvvvvi} is well-behaved as a distribution in $u$ provided $\vv\neq 0$. If $\vv=0$, a separate analysis is required, starting from the $\epsilon \neq 0$ expression.\footnote{We outline this analysis in the context of the stress tensor-contribution in the next subsection.} Similarly\footnote{There is no sum over $A$ in the following equation.}
\begin{align}
\label{tvAvBi}
{\<[T^{\vv A}(u,\vv,x_\perp),T^{\vv B}(0)]\>\over C_T} &= -{i\pi x_A x_B \over 15\vv^4} \d^{(5)} \Big(u-{x_\perp^2 \over \vv}\Big)\, ,\\
\label{tvAvAi}
{\<[T^{\vv A}(u,\vv,x_\perp),T^{\vv A}(0)]\>\over C_T} &= {i\pi \over 12\vv^3} \d^{(4)} \Big(u-{x_\perp^2 \over \vv}\Big) - {i\pi (x_A)^2 \over 15\vv^4} \d^{(5)} \Big(u-{x_\perp^2 \over \vv}\Big)\, .
\end{align}
So far we have not enforced the light sheet limit $\vv \to 0.$ However, as in the equal-time case, we see the expressions we have diverge in this limit.

In section \ref{sec:BMS}, we will integrate \eqref{tvvvvi}, \eqref{tvAvBi} and \eqref{tvAvAi} to compute commutators of light-ray operators built from components of the stress-tensor with various weight factors $\int_{-\infty}^\infty du f(u)T_{u\m}$ placed on a common light sheet. In the examples below, depending on $f(u)$, the identity contribution to commutators of these light-ray operators will turn out to be either divergent or zero.

\subsubsection{Stress-tensor contribution}
\label{sec:lsttt}

In the OPE expansion \eqref{ttope4d}, it is clear for $d=4$ that the stress-tensor contributions always come with singularities of the form $x^{-2k}$ where $k$ is an integer. 
As discussed around equation \eqref{jan1} this implies their contribution to the commutator is supported on the lightcone, and take the general form
\begin{align}
[T^{\m\n}(u,\vv,x_\perp),T^{\a\b}(u',\vv',x'_\perp)] = \sum_i f_i^{\m\n\a\b \s\r}(\p', x-x') T_{\s\r}(x') \d\left((x_\perp-x_\perp')^2 -(u-u')(\vv-\vv')\right) +\ldots\, ,\label{TT11}
\end{align}
where $f_i^{\m\n\a\b \s\r}(\p',x-x')$ are some differential operators constructed from tensor products of the vectors ${\p \over \p x'^\m}$ and $(x-x')^\m$. The ellipsis denotes the contribution from other operators and we omitted the identity contribution. We are interested in the light sheet limit $\vv \to \vv' $ of this expression. The proper procedure for this is to introduce the $i\eps$ prescription for commutators and set $\vv = \vv'$ from the outset. Integrating the delta function displayed in \eqref{TT11} against a test function $f(u',x'_\perp)$, the result has support at  $u'\rightarrow u,\, x'_\perp\rightarrow x_\perp$ in the $\epsilon\rightarrow 0$ limit. This leads to the following structure
\be
\label{scll}
\lim_{\vv \to \vv'}[T^{\m\n}(u,\vv,x_\perp),T^{\a\b}(u',\vv',x'_\perp)] = \sum_i h_i^{\m\n\a\b \s\r}(\p', x-x') T_{\s\r}(x') \d(u-u') \d(x_\perp-x_\perp')\, ,
\ee
where now the restricted differential operators $h_i^{\m\n\a\b \s\r}(\p', x-x')$ act on $(u',x_\perp').$ The OPE is an expansion in twist in the lightcone limit and the left hand side of \eqref{scll} has support only on the lightcone. This implies operators with mimimum twist contribute on the right hand side. In the absence of scalar conformal primaries with dimension $1 \leq \D \leq 2$, the identity operator and the stress tensor are the only operators with minimal twist in an interacting CFT, which is why we dropped all the other operators. When these light scalars are present they will contribute to the lightsheet commutator. In appendix \ref{app:lssusy} we discuss the situation in $\cN=1$ superconformal theories and show that these light scalars do not contribute to the commutator when there is no flavor symmetry. In the rest of this section we focus on the contribution of the stress tensor.

As in the equal-time case our method is to compute\footnote{To avoid a proliferation of definitions we use the same tensor to denote the integrated light sheet commutator and the integrated equal-time commutator (\ref{ecii}). We hope the reader does not get confused by this.}
\begin{align}
\label{gv0}
\TTTten^{[\m\n,\s\r]\a\b}_\text{int.}(w) = \int d^2x_\perp du~ \<[T^{\m\n}(u,0,x_\perp),T^{\s\r}(0)] T^{\a\b}(w)\>\, .
\end{align}
We start from the analytically continued three-point function (\ref{ttt4d}), in lightcone coordinates,
\begin{align}
\label{tttels}
\TTTten^{\m\n\s\r\a\b}_{\eps}(u,x_\perp ;w) & \equiv \TTTten^{\m\n\s\r\a\b}(u-i \eps,-i\eps,x_\perp;w) -\TTTten^{\m\n\s\r\a\b}(u+i \eps,i\eps,x_\perp;w)\, .
\end{align}
The operators are ordered by imposing Im$(w^0) >\eps >0$. We then compute \eqref{gv0} as 
\begin{align}
\label{tttelsi}
\TTTten^{[\m\n,\s\r]\a\b}_\text{int.}(w) &= \lim_{\eps\to 0} \int d^2x_\perp du~ \TTTten^{\m\n\s\r\a\b}_{\eps}(u,x_\perp ;w)\, .
\end{align}
We now present explicit examples.

\subsection*{$\boldsymbol{[T^{\vv \vv}(u,0,x_\perp),T^{\vv \vv}(0)]|_T}$}

We start with the computation of $[T^{\vv \vv}(u,0,x_\perp),T^{\vv \vv}(0)]|_T$, where the subscript $T$ refers to the stress-tensor contribution. To simplify the integrals, we set $w=(w^u,w^\vv,0).$ We perform the $u$ integral by residues and the $x_\perp$ integrals along the real lines. As before we expand
\begin{align}
\label{gvvvvvvxd}
\TTTten^{\vv\vv\vv\vv\vv\vv}_{\eps}(u,x_\perp ;w)= \sum_{n=1}^5 {\TTTten^{\vv\vv\vv\vv\vv\vv}_{\eps,n}(u,x_\perp;w)\over (x_\perp^2+i\eps(u-i\eps))^n} + \mathcal{O}(((x_\perp^2 + i\epsilon(u-i\epsilon))^0)\, .
\end{align}
The sum runs over terms that give rise to non-zero distributions in $(u,x_\perp)$ in the $\eps \to 0$ limit. Performing the computation, we find
\begin{align}
\label{gvvvvvv}
\TTTten^{[\vv\vv,\vv\vv]\vv\vv}_\text{int.}(w) = -C_T{96  i\over (w^u)^7 (w^\vv)^2}\, .
\end{align}
This is equal to
\begin{align}
\TTTten^{[\vv\vv,\vv\vv]\vv\vv}_\text{int.}(w) = -4i \p_{u} \<T^{\vv\vv}(u,0,x_2,x_3)T^{\vv\vv}(w^u,w^\vv,0,0)\>|_{u=x_2=x_3=0}\, ,
\end{align}
which implies
\begin{align}
\label{tvvvv}
[T^{\vv\vv}(u,0,x_\perp),T^{\vv\vv}(u',0,x'_\perp)]|_T = 4i\left(T^{\vv\vv}(u,0,x_\perp)+T^{\vv\vv}(u',0,x'_\perp)\right)\p_u \d(u-u')\d(x_\perp-x_\perp') + \text{td}\, .
\end{align}
where td denotes total derivative terms. Lowering the $\vv$ indices, we can rewrite \eqref{tvvvv} in terms of $T_{uu}$. It then looks exactly like the two-dimensional expression \eqref{st2d} up to the terms that are total derivatives
\begin{align}
\label{tuuuu}
[T_{uu}(u,0,x_\perp),T_{uu}(u',0,x'_\perp)]|_T = i\left(T_{uu}(u,0,x_\perp)+T_{uu}(u',0,x'_\perp)\right)\p_u \d(u-u')\d(x_\perp-x_\perp') + \text{td}\, .
\end{align}
Let us illustrate how the total derivative terms enter the commutator. The stress-tensor three-point function is linear in the three central charges, $a,b$ and $c$. In particular, we can trade $b$ for $C_T$ by writing \cite{Osborn:1993cr} 
\begin{align}
b={1\over 2}\left(14a-5c-{3C_T\over \pi^2}\right)\, .
\end{align}
With this, we decompose $\TTTten_\eps^{\vv\vv\vv\vv\vv\vv}(C_T,a,c)$ as a term linear in $C_T$ and a term linear in $a$ and $c$, but that contains no $C_T,$ as\footnote{Just for the following equation we suppress the arguments $(u,x_\perp).$ The superscripts to the left of the tensors denote the power of $C_T$.}, 
\begin{align}
\TTTten_\eps^{\vv\vv\vv\vv\vv\vv}(C_T,a,c) = {}^1\TTTten_\eps^{\vv\vv\vv\vv\vv\vv}C_T+ {}^0\TTTten_\eps^{\vv\vv\vv\vv\vv\vv}(a,c)\, .
\end{align}
The term linear in $C_T$ gives the result \eqref{gvvvvvv}, while ${}^0\TTTten_\eps^{\vv\vv\vv\vv\vv\vv}(a,c)$ gives the total derivative terms, that is,
\be \int du d^2x_\perp {}^0\TTTten_\eps^{\vv\vv\vv\vv\vv\vv}(u,x_\perp;w)=0\, .
\ee
To show ${}^0\TTTten_\eps^{\vv\vv\vv\vv\vv\vv}(a,c)$ is non-zero as a distribution in $(u,x_\perp)$ we compute its second moment in $u$ 
\begin{align}
\label{gvvvvvvu2}
\lim_{\eps\to 0} \int du\, u^2 \int d^2x_\perp  {}^0\TTTten^{\vv\vv\vv\vv\vv\vv}_{\eps}(u,x_\perp;w) = -\frac{3 i \pi ^3 (9969773439
	a-3304074649
	c)}{2097152
	(w^u)^5 (w^{\vv})^2}\, ,
\end{align}
while its first moment vanishes
\begin{align}
\label{uintvv}
\int du\, u \int d^2x_\perp  {}^0\TTTten^{\vv\vv\vv\vv\vv\vv}_{\eps}(u,x_\perp;w) =0\, .
\end{align}
This will turn out to be important in computing the action of $\cK(x_\perp)=\int_{-\infty}^{\infty} du\, u T_{uu}(u,0,x_\perp)$ on various light integrals of the stress-tensor in section \ref{sec:BMS}.

We delegate the computation of $[T^{\vv A},T^{\vv \vv}]$ and $[T^{\vv A},T^{\vv B}]$ to appendix \ref{app:tttlight} and here we quote the results
\begin{align}
\label{tuAuu}
[T_{u A}(u,0,x_\perp),T_{uu}(u',0,x'_\perp)]|_T &= iT_{uu}(u,0,x_\perp) \d(u-u') \p_A \d(x_\perp-x_\perp') + \text{td}\, ,\\
\label{tuAuB}
[T_{u A}(u,0,x_\perp),T_{u B}(u',0,x'_\perp)]|_T&= i\left(T_{u B}(u,0,x_\perp)\p_A +T_{u A}(u',0,x'_\perp)\p_B\right)  \d(x_\perp-x_\perp')\d(u-u') + \text{td}\, .
\end{align}

\section{BMS and Virasoro algebras in CFT$_4$ \label{sec:BMS}}

In this section we compute integrals of stress-tensor commutators on a common light sheet that we derived in the previous section for CFT$_4.$ First we derive the BMS algebra of \cite{Cordova:2018ygx} for generic interacting CFTs. We then define the four-dimensional analogues of the Virasoro generators \eqref{ln2d}.  We further discuss their three-point functions with scalar operators and the difficulties involved in computing their commutators.

\subsection{BMS}
\label{subsec:BMS}

Following \cite{Cordova:2018ygx}, we define the following light-ray operators built from components of the stress-tensor
\begin{align}
\cE(\vv,x_\perp)&=\int_{-\infty}^{\infty} du T_{uu}(u,\vv,x_\perp),\\
\cK(\vv,x_\perp)&=\int_{-\infty}^{\infty} du \,u T_{uu}(u,\vv,x_\perp),\\
\cN_A(\vv,x_\perp)&=\int_{-\infty}^{\infty} du T_{uA}(u,\vv,x_\perp).
\end{align}
We compute the algebra they satisfy on the light sheet $\vv=0$ by computing integrals of the local stress-tensor commutators that we computed in section \ref{STcomm4d}. When they are placed on the same light sheet  at $\vv=0$, we drop the first argument and denote, for example, $\cE(\vv=0,x_\perp) \equiv \cE(x_\perp).$  

As discussed in the Introduction only the identity and the stress-tensor contribute to commutators in the light sheet limit, which is the regime we discuss in the following. We start by deriving the contribution of the identity operator. 

\subsubsection{Identity contribution}

We start with $[\cK,\cE].$ Using (\ref{tvvvvi}), we obtain 
\begin{align}
\left.[\cK(\vv,x_\perp),\cE(0)]\right|_{id} = -{i\pi C_T\over 15\vv^2} \int_{-\infty}^{\infty} du\,u \int_{-\infty}^{\infty} du'  \d^{(5)} \Big(u-u' -{x_\perp^2 \over \vv}\Big)=0\, .
\end{align}
In fact, the commutators (\ref{tvvvvi}),(\ref{tvAvBi}),(\ref{tvAvAi}) all involve at least fourth derivatives of the delta function. This implies the identity contribution to commutators of $\cE,\cK,\cN_A$ among themselves vanishes. We now discuss the stress-tensor contribution to these commutators.

\subsubsection{Stress-tensor contribution}

For the commutators among $\cE$ and $\cK$ we need (\ref{tuuuu}) which we remind here for convenience
\begin{align}
[T_{uu}(u,0,x_\perp),T_{uu}(u',0,x'_\perp)]|_T = i\left(T_{uu}(u,0,x_\perp)+T_{uu}(u',0,x'_\perp)\right)\p_u \d(u-u')\d(x_\perp-x_\perp')\, .
\end{align}
We dropped total derivatives since they vanish upon integration. To justify this for commutators involving $\cK$ below it is important that the total derivative contributions have first $u$-moment zero, which is what we showed in equation \eqref{uintvv}. The absence of derivatives on the transverse delta functions implies that we can freely exchange $x_\perp$ with $x'_\perp.$ 

We start with the commutator of two ANEC operators.

\subsection*{$\boldsymbol{[\cE(x_\perp),\cE(x'_\perp)]}$}

To compute the commutator of two ANEC operators, we plug in 
\begin{align}
[\cE(x_\perp), \cE(x'_\perp)] & =i \d(x_\perp-x_\perp')\int_{-\infty}^{\infty}du \int_{-\infty}^{\infty}du' \left(T_{uu}(u,0,x_\perp)+T_{uu}(u',0,x'_\perp)\right)\p_u \d(u-u')\nonumber\\
&=0\, ,
\end{align}
where for the first term we performed the $u'$ integral first using $\p_u \d(u-u')=-\p_{u'} \d(u-u')$ and in the second term we first performed the $u$ integral. We thus showed that our method reproduces the fact that ANEC operators on a common light sheet commute, which was proven in \cite{Kologlu:2019bco}. Let us move to a case where the commutator is non-vanishing.

\subsection*{$\boldsymbol{[\cK(x_\perp),\cE(x'_\perp)]}$}

We want to compute 
\begin{align}
[\cK(x_\perp),\cE(x'_\perp)] &=i \d(x_\perp-x_\perp')\int_{-\infty}^{\infty}du \,u \int_{-\infty}^{\infty}du' \left(T_{uu}(u,0,x_\perp)+T_{uu}(u',0,x'_\perp)\right)\p_u \d(u-u')\no\\
&\equiv i\d(x_\perp-x_\perp')\cC_{\cK\cE}(x'_\perp)\, .
\end{align}
The first term contributes to $\cC_{\cK\cE}(x'_\perp)$ as
\begin{align}
- \int_{-\infty}^{\infty}du'\, T_{uu}(u',0,x_\perp) - \int_{-\infty}^{\infty}du'\,u'   \p_{u'} T_{uu}(u',0,x_\perp)\, ,
\end{align}
while the second contributes as 
\begin{align}
- \int_{-\infty}^{\infty}du\,u \int_{-\infty}^{\infty}du'\, T_{uu}(u',0,x'_\perp) \p_{u'} \d(u-u')=\int_{-\infty}^{\infty}du\,u   \p_{u} T_{uu}(u,0,x'_\perp)\, .
\end{align}
Adding the two, we get, 
\begin{align}
\label{kec}
[\cK(x_\perp),\cE(x'_\perp)] &=-i \d(x_\perp-x_\perp')\cE(x'_\perp)\, .
\end{align}
This commutator was derived in \cite{Casini:2017vbe,Cordova:2018ygx,belin2020stress}.

\subsection*{$\boldsymbol{[\cN_{A}(x_\perp), \cE(x'_\perp)]}$}

For the commutator $[\cN_{A}(x_\perp), \cE(x'_\perp)]$ we need (\ref{tuAuu}), which we remind here for convenience
\begin{align}
[T_{u A}(u,0,x_\perp),T_{uu}(u',0,x'_\perp)]|_T = iT_{uu}(u,0,x_\perp) \d(u-u') \p_A \d(x_\perp-x_\perp')\, .
\end{align}
To drop the total derivative terms which could in principle contribute to commutators involving $\cK(x_\perp)$ we used (\ref{uintvA}). The commutator is then given as 
\begin{align}
[\cN_{A}(x_\perp), \cE(x'_\perp)] &=i\int_{-\infty}^{\infty}du \int_{-\infty}^{\infty}du' T_{uu}(u,0,x_\perp) \d(u-u')\p_A \d(x_\perp-x_\perp')\no\\
&= -i \cE(x_\perp) \p_{A'} \d(x_\perp-x_\perp')\,,
\end{align}
where $\p_{A'}$ denotes a derivative acting on $x'_\perp.$ We can express this result at $x'_\perp$ by adding $i\p_{A'}\left[\d(x_\perp-x_\perp')\left( \cE(x_\perp)-\cE(x'_\perp) \right)\right]=0,$ which yields,
\begin{align}
[\cN_{A}(x_\perp), \cE(x'_\perp)] &=-i \d(x_\perp-x_\perp') \p_{A'}\cE(x'_\perp) + i \cE(x'_\perp)\p_{A} \d(x_\perp-x_\perp')\, .
\end{align}
We delegate the computation of the rest of the commutators to appendix \ref{app:BMS}. The result is
\begin{align*}
[\cE(x_\perp),\cE(x'_\perp)]&=0\, ,\\
[\cK(x_\perp),\cK(x'_\perp)]&=0\, ,\\
[\cK(x_\perp),\cE(x'_\perp)] &=-i \d(x_\perp-x_\perp')\cE(x'_\perp)\, ,\\
[\cN_{A}(x_\perp), \cE(x'_\perp)] &=-i \d(x_\perp-x_\perp') \p_{A'}\cE(x'_\perp) + i \cE(x'_\perp)\p_{A} \d(x_\perp-x_\perp')\, ,\\
[\cN_{A}(x_\perp), \cK(x'_\perp)] &=-i \d(x_\perp-x_\perp') \p_{A'}\cK(x'_\perp) + i \cK(x'_\perp)\p_{A} \d(x_\perp-x_\perp')\, ,\\
[\cN_{A}(x_\perp), \cN_{B}(x'_\perp)] &= -i \d(x_\perp-x_\perp')\p_{A'} \cN_{B}(x'_\perp) +i \cN_{B}(x'_\perp)\p_{A} \d(x_\perp-x_\perp') +i \cN_{A}(x'_\perp)\p_{B} \d(x_\perp-x_\perp')\, ,
\end{align*}
as argued in \cite{Cordova:2018ygx}.

\subsection{An infinite set of generalizations of the ANEC operator}
\label{sec:4dvirint}

In CFT$_2$ we have seen how an infinite set of light-ray operators built from the stress-tensor (\ref{ln2d}) act on scalar fields and satisfy the Virasoro algebra. Inspired by this we define their $d=4$ generalization. The first property different from the $d=2$ case is the dependence of the stress tensor $T_{uu}$ on all coordinates $(u,\vv,x_\perp).$ This implies, just like the BMS generators discussed above, the 4d ``Virasoro''s will depend of $(\vv,x_\perp),$ after integrating them over the light ray $(\vv=\text{constant},x_\perp=\text{constant}).$ The most general candidate is then $\int du f(u) T_{uu}(u,\vv,x_\perp)$ where $f(u) \sim u^{4}$ for large $u;$ this property is required for these operators to have well defined matrix elements. Just as in the $d=2$ case demanding these annihilate the right or left vacuum in Wightman three point functions with two scalar primaries and Wightman two point functions with the arbitrary components of the stress tensor we arrive at the following definition
\begin{align}
\mathcal{L}_n^{\text{gen}}(\vv, x_\perp)=R^{1-\a-\b-\g}\int_{-\infty}^{\infty}du (iR+u)^{-n+\a} (iR-u)^{n+\b} (s-u)^\g T_{uu}(u,\vv ,x_\perp)\, .
\end{align}
where $R,s \in \mathbb{R}_{>0}.$ These operators have mass dimension 2, and have finite correlators for $\a+\b+\g=4.$ We further demand $(\mathcal{L}_n^{\text{gen}})^\dagger =\mathcal{L}_{-n}^\text{gen}$ and this sets $\a=\b.$ As we will show shortly, insisting these operators are closed under a collinear subalgebra of the conformal algebra that leaves invariant the light ray $(\vv,x_\perp)$ singles out
\begin{align}
\label{ceav}
\mathcal{L}_n(\vv, x_\perp)\equiv R^{-3}\int_{-\infty}^{\infty}du (iR+u)^{-n+2} (iR-u)^{n+2} T_{uu}(u,\vv,x_\perp)\, .
\end{align}
where $R \in \mathbb{R}_{>0}.$ To reiterate, these operators have mass dimension 2 and satisfy $\cL_n^\dagger = \cL_{-n}.$

\subsubsection{Collinear subalgebra}

We first discuss the action of the collinear subalgebra of the conformal algebra on these operators following \cite{Braun:2003rp,belin2020stress}. The collinear subalgebra is defined as the subalgebra of the full conformal algebra that leaves the light-ray $\vv=0,~x_\perp=0$ invariant, and is generated by\footnote{$D,M_{\m\n},K_\a$ denote respectively dilations, Lorentz tranformations, and special conformal transformations.} 
\begin{align}
J_{-1} \equiv iP_u,\quad ~~~J_0={i\over 2}(D-M_{\vv u})\quad ,~~~J_1=-iK_\vv \, ,
\end{align}
and
\begin{align}
\bar{J}_0={i\over 2}(D-M_{\vv u})\, .
\end{align}
which commutes with $J_{-1},J_0,J_1.$ These generators satisfy an $SL(2,\mathbb{R})$ algebra
\begin{align}
[J_0,J_{\pm 1}] = \mp J_{\pm 1},\qquad ~~~[J_1,J_{- 1}] = 2 J_0\, .
\end{align}
Their action on the stress-tensor at $x_\perp =0$ is given by
\begin{equation}
\begin{aligned}
\label{coll}   
[J_{-1}, T_{uu}(x)]&= \p_uT_{uu}(x)\, ,\\
[J_{0}, T_{uu}(x)] &=  \left(3+u\p_u\right)T_{uu}(x)\, ,\\
[J_{1}, T_{uu}(x)] &=  \left(6u+u^2\p_u\right)T_{uu}(x)\, ,
\end{aligned}
\end{equation}
where $x=(u,\vv,0).$ Integrating (\ref{coll}) along the light-ray $\vv =0,~x_\perp=0$, we obtain 
\begin{align}
[J_{-1}, \mathcal{L}_n] &= -{i\over R}\left( n\mathcal{L}_n+ {(n-2)\over 2} \mathcal{L}_{n+1} +{(n+2)\over 2}\mathcal{L}_{n-1} \right)\, ,\\
[J_{0}, \mathcal{L}_n] &= {(2-n)\over 2} \mathcal{L}_{n+1} +{(n+2)\over 2}\mathcal{L}_{n-1}\, ,\\
[J_{1}, \mathcal{L}_n] &=iR\left( -n\mathcal{L}_n+ {(n-2)\over 2} \mathcal{L}_{n+1} +{(n+2)\over 2}\mathcal{L}_{n-1} \right)\, .
\end{align}
We see that the set (\ref{ceav}) is closed under the action of the collinear subalgebra.

There exists a specific subset of the whole family of operators defined in (\ref{ceav}) that is also closed under the action of the collinear algebra, and they are given by 
\begin{align}
G\equiv\{\cL_{-2},\cL_{-1},\cL_{0},\cL_{1},\cL_{2}\}\, .\label{Virg4def}
\end{align}
Here $G$ stands for \textit{global} in analogy with the two-dimensional case. To see this we note that generators in $G$ are a linear combination of 
\begin{align}
\label{ltk4}
\tilde{\cL}_k \equiv \int du\, u^{k+2}T_{uu},\quad ~~~k=-2,\, -1,\, 0,\, 1,\, 2\, .
\end{align}
The ANEC operator $\cE$ corresponds to $k=-2$ in the notation of \eqref{ltk4} and is given as a linear combination
\begin{align}
\cE={3\over 8R}\mathcal{L}_0 + {1\over 4R}(\mathcal{L}_1+\mathcal{L}_{-1}) + {1\over 16R}(\mathcal{L}_2+\mathcal{L}_{-2})\, .
\end{align}
It is annihilated by $J_{-1}$ and is an eigenoperator of $J_0$ with eigenvalue 2
\begin{align}
[J_{-1},\cE(\vv,0)]=0,\quad ~~~[J_0,\cE(\vv,0)]=2 \cE(\vv,0).
\end{align} 
Acting on these five operators with $J_1$ we get\footnote{In the following the light-ray operators are understood to be localized at $x_\perp =0.$}
\begin{align}
[J_1,\cE]\sim \tilde{\cL}_{-1},~~[J_1,\tilde{\cL}_{-1}]\sim \tilde{\cL}_0,~~[J_1,\tilde{\cL}_0]\sim \tilde{\cL}_1,~~[J_1,\tilde{\cL}_1]\sim \tilde{\cL}_2,~~[J_1,\tilde{\cL}_2]=0\, .
\end{align} 
Thus, the five operators \eqref{Virg4def} in $G$ form a 5-dimensional representation of $SL(2,\mathbb{R})$.

\subsubsection{Correlators of light-ray operators $\cL_n$}

We continue with the computation of correlators of (\ref{ceav}). We start with the three-point function, where the operators are not ordered, that is given as  \cite{Osborn:1993cr}
\begin{align}
\< T^{\vv \vv}(u,0,0)\mathcal{O}(u_1,\vv,x_{1,\perp}) \mathcal{O}(u_2,-\vv,x_{2,\perp})\>&={C_{TOO}\vv^2 (\vv(u_2-u_1)+x_{1,\perp}^2+x_{2,\perp}^2)^2\over x_{12}^{2\D-2}((u-u_1)\vv+x_{1,\perp}^2)^{3} ((u_2-u)\vv+x_{2,\perp}^2)^{3} }\, ,
\end{align}
where $x_{12}^2 = 2(u_2-u_1)\vv + (x_{1,\perp}-x_{2,\perp})^2$. We integrate this with various weight factors and as usual the $i\eps$ prescription is crucial in enforcing the various orderings $\<TOO\>,\<OTO\>,\<OOT\>.$ If we ignore the necessary $i\epsilon$ that are necessary for the appropriate ordering, the (unordered) integral that we need to perform is given by 
\begin{align}
&\braket{\mathcal{L}_n(0)\mathcal{O}(u_1,\vv,x_{1,\perp})\mathcal{O}(u_2,-\vv,x_{2,\perp})}\nonumber\\
&\qquad ={C_{TOO}\over x_{12}^{2\D-2}} \vv^2 (\vv(u_2-u_1)+x_{1,\perp}^2+x_{2,\perp}^2)^2\int_{-\infty}^\infty du {(iR+u)^{-n+2} (iR-u)^{n+2}\over ((u-u_1)\vv+x_{1,\perp}^2)^{3} ((u_2-u)\vv+x_{2,\perp}^2)^{3} }\, . \label{greg2}
\end{align}
We perform this integral with residues. The integrand dies off along the positive and negative imaginary axes so we can close the contour on either side. We analyze the singularity structure of the integrand. The Wightman function has two poles, which are localized at positions 
\begin{align}
\label{lnoop}
u\rightarrow u_1 - {x_{1,\perp}^2\over \vv}\equiv  \tilde{u}_1,~~~\qquad u\rightarrow  u_2 + {x_{2,\perp}^2\over \vv}\equiv \tilde{u}_2\, .
\end{align}
Pushing these singularities slightly along the imaginary axes precisely corresponds to enforcing a relative ordering of operators inside the correlator \eqref{greg2}. Depending on the value of $n,$ more poles are introduced at $u=iR$ and $u=-iR.$ 

The subset corresponding to the global subalgebra $G$, which are the operators $\mathcal{L}_n$ with $n=-2,-1,0,1,2$ is special. The only poles are those seen in (\ref{lnoop}), and for these special values of $n$, we obtain 
\be 
\label{glo4doo}
\braket{\mathcal{L}_n(0)\mathcal{O}(u_1,\vv,x_{1,\perp})\mathcal{O}(u_2,-\vv,x_{2,\perp})} = \braket{\mathcal{O}(u_1,\vv,x_{1,\perp})\mathcal{O}(u_2,-\vv,x_{2,\perp})\mathcal{L}_n(0)} = 0\, ,
\ee
exactly analogous to the two-dimensional case.

For $n\geq 3$ the integrand has an additional pole at $u=-iR$ and for $n\leq -3$ it has an additional pole at $u=iR.$ Performing the integrals with the appropriate $i\epsilon$ prescription yields 
\begin{align}
\braket{\mathcal{O}(u_1,\vv,x_{1,\perp})\mathcal{O}(u_2,-\vv,x_{2,\perp})\mathcal{L}_n(0)} &=0\, ,\qquad \text{for } n\geq 3\, ,\\
\braket{\mathcal{L}_n(0)\mathcal{O}(u_1,\vv,x_{1,\perp})\mathcal{O}(u_2,-\vv,x_{2,\perp})} &=0\, ,\qquad \text{for } n\leq -3\,.
\end{align}
Combined with (\ref{glo4doo}) this implies
\begin{align}
\mathcal{L}_n|0\>=0,~~n\geq -2\, .
\end{align}
We present further evidence for this claim by integrating stress-tensor two-point functions in appendix \ref{app:itt}. Computing arbitrary three-point functions of $\cL_n$ with two scalar primaries for a given $n$ is a straightforward exercise. The answer is always finite, and we do not have to regularize the operators $\mathcal{L}_n$.

\subsubsection{The algebra $[\cL_m,\cL_n]$}

In this subsection we make some remarks on the algebra the light-ray operators (\ref{ceav}) satisfy, when placed on a common light sheet. The central term is computed by integrating (\ref{tvvvvi}). We define
\be 
\label{cmn4dm}
c_{m,n} = {\<[\mathcal{L}_m(\vv, x_\perp), \mathcal{L}_n(0)]\> \over C_T}\, ,
\ee
and discuss its general features in appendix \ref{sec:vir4id}. The central term (\ref{cmn4dm}) turns out to be a function of the combination $R\vv,$ $c_{m,n}=c_{m,n}(R\vv,x_\perp).$ As we discuss in appendix \ref{sec:vir4id} it simplifies in the light sheet limit $R\vv \to 0$ and takes the form 
\begin{align}
\label{v4dct}
\lim_{R\vv \to 0} c_{m,n}(R\vv , x_\perp) = {2\pi^2\over 75 }|m|(m^4-5m^2+4)\d_{m+n,0} \left({m\pi \over 12}\D_\perp\d(x_\perp) - \lim_{R\vv  \to 0} \left[ {2i\pi \over R\vv } \d(x_\perp) \right]\right) \, .
\end{align}
where $\d(x_\perp)=\d(x_2)\d(x_3)$ denotes the transverse delta function and $\D_\perp =\p_2^2+\p_3^2$ denotes the transverse Laplacian. As expected, the central contribution to the global subalgebra (\ref{Virg4def}) vanishes, and non-vanishing values of the central term are UV divergent.

Non-identity operator contributions to the commutator $[\cL_m(\vv,x_\perp),\cL_n(0)]$ are more subtle, even in the light sheet limit $\vv \to 0$ where only the stress-tensor contributes. Computation of the stress-tensor contribution follows by integrating the local commutator (\ref{tuuuu}), however, without the knowledge of total derivative terms in the local stress-tensor commutator we cannot compute the stress-tensor contribution to $[\cL_m(0,x_\perp),\cL_n(0)]$ fully. In appendix \ref{app:vir4dst} we perform the computation neglecting total derivative terms and discuss the connection to a related discussion in \cite{Casini:2017vbe}.

We conclude this section with comments on the divergencies we encounter in our commutators, namely \eqref{v4dct}. The first term in the brackets contains a delta function, which is standard in commutators in QFT when the two operators are forced to lie on a hyperplane such as the equal time limit. The usual prescription to make sense of this is to smear the operators along the directions forced to a limit by the delta function, namely transverse directions. The infinity $\lim_{\vv  \to 0}{1\over \vv}$ in the second term is a bit more subtle. One way to address this is to keep the two operators slightly displaced along the $\vv$ direction. We leave an investigation of the implications of this prescription on possible applications of the symmetry algebra we uncovered to future work.

\section{Discussion and future directions}
\label{sec:conc}

In this work we discussed commutators of local operators in CFT, focusing primarily on the stress-tensor. Using a careful distributional analysis 
of the OPE, we computed commutators using the $i\eps$ prescription and showed that this gives rise to the standard commutators of free field theory 
and of the Virasoro algebra in CFT$_2$. 
In four-dimensional conformal field theories, we do not have control over the full $TT$ OPE, and therefore cannot compute commutators
in full generality. In two special cases, equal-time and light sheet commutators, certain simplifications occur which allow us to do more
explicit computations. In particular, we are able to derive the equal-time commutators that give rise to the Poincar\'e algebra using the known
structure of the stress-tensor three-point function.

A major simplification that arises in the light sheet limit is that the OPE becomes an expansion in twist, which 
truncates the commutator to the contribution of the identity operator and the stress-tensor. 
Integrating these local commutators with relevant weight factors we were able to derive the BMS algebra proposed in \cite{Cordova:2018ygx}. 
We then defined light-ray operators $\int du (iR+u)^{2-n} (iR-u)^{2+n} T_{uu}$ in four-dimensional CFTs, inspired by their 2d analogues that 
give rise to the well known Virasoro algebra. In CFT$_4$ we showed these are closed under an $SL(2,\mathbb{R})$ subalgebra of the conformal algebra that 
preserves the light-ray, obey nice Hermiticity relations, and possess a semi-infinite subset that annihilates the vacuum. We made some initial steps in computing the algebra that these new operators satisfy. We extracted the central term from the stress-tensor 
two-point function $\<[T,T]\>$, and the stress-tensor contribution from the three-point function $\<[T,T]T\>$, and sketched what remains to 
be done in order to perform a more complete analysis.

Rather than using the $TT$ OPE directly, we obtained several results indirectly by employing the stress-tensor three-point function.
It is conceivable that certain aspects of commutators are more easily extracted by using the conformal $T$ block in the $TT$ OPE instead. 
The $T$ block in the $TT$ OPE and the $TTT$ three-point functions contain equivalent information, and can be connected using e.g. the
formalism of shadow operators \cite{dolan2012conformal,Simmons_Duffin_2014,Czech_2016,de_Boer_2016,Haehl_2019}. Unfortunately, we have been unable to use this relation to our advantage. It would be interesting to explore this connection further.  

There are several subtleties and issues which merit a further investigation. Light-ray operators of the type that we considered are 
in general unbounded operators which need not be defined on arbitrary states. It is therefore not obvious that their product or
commutator generically make sense \cite{Kologlu:2019mfz}. A second and perhaps not unrelated point is that we extracted the 
commutators using the (analytically)
continued OPE. Convergence of the Lorentzian OPE is a subtle issue and it is not a priori clear whether one is actually allowed to use the 
OPE when computing the commutator inside general correlation functions. In fact there is a disagreement between a holographic 
computation and the field theory computation of the BMS commutator $[{\cal K},{\cal E}]$ \cite{belin2020stress}, which could be a large $N$
artifact but also a signal that there is an issue with the precise definition of these operators. 

If light-ray operators are strictly speaking ill-defined, one could try to modify their definition to make them well-defined, 
but it is not clear what such a modification could be. One suggestion is to add an explicit regulator \cite{belin2020stress}
so that the large $u$ behavior is improved, and which has the effect that one can generically close contours in the $u$-plane as long as the 
$u$-behavior is analytic and there are no branch cuts. Alternatively, one could think more seriously about the conformal embedding
of Minkowski space in the Lorentzian cylinder and only consider states on Minkowski spacetime which admit a proper extension to
the full cylinder. Given the discreteness of the spectrum of the theory on the cylinder this might also alleviate some of the problematic
features of the theory on the Minkowski plane. It would be interesting to investigate this in more detail.

Finally, an interesting future direction is the exploration of the implications of the symmetry-like structures that 
emerge on the light sheet. In CFT$_2,$ Virasoro symmetry leads to powerful constraints on the structure of correlation functions and 
consequently on the data of the CFT. It would be very interesting to study whether analogous constraints, in the form of Ward identities 
and sum rules, are implied by their four-dimensional cousins. Finally, another interesting question is whether these operators have interesting implications when placed at future null infinity  (for a review of some recent work, see \cite{strominger2018lectures}).

\acknowledgments

It is a pleasure to thank Tarek Anous, Jackson Fliss and Austin Joyce for helpful discussions. We also want to specifically thank Diego Hofman for initial collaboration and multiple insightful discussions throughout the project. JdB is supported by the European Research Council
under the European Unions Seventh Framework Programme (FP7/2007-2013), ERC Grant agreement ADG
834878.
M.B. and G.M.\ are supported in part by the ERC starting grant {\scriptsize{GENGEOHOL}} (grant agreement No 715656).

\appendix

\section{Details of two-dimensional computations}
\label{app:vir2}

\subsection{Stress-tensor commutator}

In this appendix, our goal is to compute the holomorphic stress-tensor commutator starting from the stress-tensor OPE
\begin{align}
\label{lope2}
T_{\vv\vv}(\vv)T_{\vv\vv}(\vv') =  {c\over 8\pi^2(\vv-\vv')^4} - {T_{\vv\vv}(\vv')\over \pi(\vv-\vv')^2} -{\p_{\vv'} T_{\vv\vv}(\vv') \over 2\pi(\vv-\vv')}\, ,
\end{align}
where we dropped the regular terms since they do not contribute to the commutator.

The central term is computed in section \ref{sec:stcft2} and here we compute the stress-tensor contribution. For the contribution of the middle term in (\ref{lope2}) we define
\begin{align}
g^{\mp}_{1,\eps} (\vv,\vv') = - {T_{\vv\vv}(\vv')\over \pi(\vv-\vv'\mp i\eps)^2}\, ,
\end{align}
We define the commutator $g^c_{1} (\vv,\vv')$ as a distribution in $\vv,\vv'$. Integrating against an arbitrary test function $h(\vv')$
\begin{align}
\int_{-\infty}^\infty d\vv'\, g^c_{1} (\vv,\vv')h(\vv') = -{1\over \pi} \lim_{\eps\to 0} \int_{-\infty}^\infty  d\vv' \left[ {T_{\vv\vv}(\vv') h(\vv')\over (\vv-\vv'- i\eps)^2} - {T_{\vv\vv}(\vv') h(\vv')\over (\vv-\vv'+ i\eps)^2}  \right]\, .
\end{align}
Performing the $\vv'$ integral with residues and taking the $\eps \to 0$ limit yields 
\begin{align}
\label{mt2d}
\int_{-\infty}^\infty d\vv'\, g^c_{1} (\vv,\vv')h(\vv')  = 2i \left(T_{\vv\vv}(\vv) \p_\vv h(\vv) + h(\vv) \p_\vv T_{\vv\vv}(\vv)  \right)\, .
\end{align}
For the contribution of the last term in (\ref{lope2}) we define
\begin{align}
g^{\mp}_{2,\eps} (\vv,\vv') = - { \p_{\vv'}T_{\vv\vv}(\vv')\over 2\pi (\vv-\vv'\mp i\eps)}\, ,
\end{align}
and a similar computation gives
\begin{align}
\label{lt2d}
\int_{-\infty}^\infty d\vv'\, g^c_{2} (\vv,\vv')h(\vv')  = -i h(\vv) \p_\vv T_{\vv\vv}(\vv) \, .
\end{align}
Combining (\ref{mt2d}) and (\ref{lt2d}), we get 
\begin{align}
\int_{-\infty}^\infty d\vv' \left(g^c_{1}(\vv,\vv')+g^c_{2}(\vv,\vv')\right)h(\vv') = i \int_{-\infty}^\infty d\vv' \d(\vv-\vv') \left(2T_{\vv\vv}(\vv') \p_{\vv'} h(\vv') + h(\vv') \p_{\vv'} T_{\vv\vv}(\vv') \right)\, ,
\end{align}
which we integrate by parts to get 
\begin{align}
g^c_{1}(\vv,\vv')+g^c_{2}(\vv,\vv') &= -2iT_{\vv\vv}(\vv') \p_{\vv'} \d(\vv-\vv') -i \d(\vv-\vv') \p_{\vv'} T_{\vv\vv}(\vv')\, .
\end{align}
Combining this with (\ref{g0c}) and adding $i\p_{\vv'}\left[ \d(\vv-\vv')\left( T_{\vv\vv}(\vv') - T_{\vv\vv}(\vv) \right) \right]=0$, we finally arrive at the usual form of the Virasoro algebra written as a commutator, which is given by 
\begin{align}
\label{2dstca}
[T_{\vv\vv}(\vv), T_{\vv\vv}(\vv')] = i\left(T_{\vv\vv}(\vv) + T_{\vv\vv}(\vv')\right) \p_\vv \d(\vv-\vv')  -{ic\over 24 \pi} \p^3_\vv \d(\vv-\vv')\, .
\end{align}
This shows how we can extract commutators from the OPE. 

\subsection{Virasoro algebra}

In this appendix, our goal is to compute explicitly commutators of the new set of two-dimensional generators (\ref{ln2d}) that we introduced in the main text. Their definition is remembered here for convenience 
\begin{align}
L_n = -{1\over 2R} \int_{-\infty}^{\infty}d\vv (iR+\vv)^{1-n} (iR-\vv)^{n+1} T_{\vv\vv}(\vv)\, .
\end{align}
We want to show commutators of these operators produce the Virasoro algebra. The commutator is obtained by integrating the stress-tensor commutator (\ref{2dstca}). We start with the central term
\begin{align}
\left.[L_m, L_n]\right|_{id} &= -{ic\over 96 \pi R^2} \int_{-\infty}^{\infty}d\vv (iR+\vv)^{1-m} (iR-\vv)^{m+1} \int_{-\infty}^{\infty}d\vv' (iR+\vv')^{1-n} (iR-\vv')^{n+1} \p^3_\vv \d(\vv-\vv')\no
\\&={m(m^2-1)c\over 12 \pi} {\sin[ (m+n)\pi]\over (m+n)} \no
\\
\label{centr}
&= {(m^3-m)c\over 12}\d_{m+n,0}\, .
\end{align}
The stress-tensor contribution is given by 
\begin{align}
\left.[L_m, L_n]\right|_{T} &= {i\over 4R^2} \int_{-\infty}^{\infty}d\vv \int_{-\infty}^{\infty}d\vv' \m_m(\vv) \m_n(\vv') \left( T_{\vv\vv}(\vv) + T_{\vv\vv}(\vv') \right) \p_\vv \d(\vv-\vv')\, ,\label{eqapp:T}
\end{align}
where we have defined $\mu_m(\vv) = (iR + \vv)^{1-m} (iR-\vv)^{m+1}$.
We compute the first term of \eqref{eqapp:T} by first performing the $\vv$ integral by parts\footnote{Using delta function identities the answer can be shown to be independent of the order of integration. The boundary terms vanish, and this is the case throughout the paper whenever we perform an integration by parts.} and we get 
\begin{align}
\left.[L_m, L_n]\right|^1_T = {i\over 4R^2} \int_{-\infty}^{\infty}d\vv' (iR+\vv')^{1-m-n} (iR-\vv')^{1+m+n} \left[ (2(iRm+\vv') T_{\vv\vv}(\vv') + (R^2+\vv'^2) \p_{\vv'} T_{\vv\vv}(\vv') \right]\, ,\label{eq:app1}
\end{align}
and the second term of \eqref{eqapp:T} by first performing the $\vv'$ integral by parts to get 
\begin{align}
\left.[L_m, L_n]\right|^2_T=-{i\over 4R^2} \int_{-\infty}^{\infty}d\vv (iR+\vv)^{1-m-n} (iR-\vv)^{1+m+n} \left[  (2(iRn+\vv) T_{\vv\vv}(\vv) + (R^2+\vv^2) \p_{\vv} T_{\vv\vv}(\vv) \right]\, .\label{eq:app2}
\end{align}
Adding \eqref{eq:app1}, \eqref{eq:app2} and \eqref{centr}, we obtain the usual Virasoro algebra
\begin{align}
\label{lmlnntd}
[L_m, L_n] = (m-n) L_{m+n} +{c\over 12}(m^3-m)\d_{m+n,0}\, .
\end{align}

\section{Stress-tensor commutators in the light sheet limit}
\label{app:tttlight}

In this appendix we compute $[T^{\vv A},T^{\vv \vv}]$ and  $[T^{\vv A},T^{\vv B}]$ in CFT$_4$ in the light sheet limit. We follow the discussion in section \ref{sec:lsttt}.

\subsection*{$\boldsymbol{[T^{\vv A}(u,0,x_\perp),T^{\vv \vv}(0)]|_T}$}

Setting $w=(0,w^\vv,0,w_3)$ we define $\TTTten_\eps^{\vv3\vv\vv\vv\vv}$ following (\ref{tttels}) and compute
\begin{align}
\TTTten^{[\vv 3,\vv\vv]\vv\vv}_\text{int.}(w) \equiv  \int d^2x_\perp du~ \<[T^{\vv 3}(u,0,x_\perp),T^{\vv \vv}(0)] T^{\vv\vv}(w)\>\, ,
\end{align}
using (\ref{tttelsi}). We find
\begin{align}
\TTTten^{[\vv 3,\vv\vv]\vv\vv}_\text{int.}(w) = C_T{96i (w^\vv)^4 \over  (w_3)^{13}} = 2i  \p_{x_3} \<T^{\vv\vv}(u,0,x_2,x_3)T^{\vv\vv}(0,w^\vv,0,w_3)\>|_{u=x_2=x_3=0}\, .
\end{align}
This implies
\begin{align}
[T_{u A}(u,0,x_\perp),T_{uu}(u',0,x'_\perp)]|_T = iT_{uu}(u,0,x_\perp) \d(u-u') \p_A \d(x_\perp-x_\perp') + \text{td}\, .
\end{align}
In section \ref{sec:BMS} where we reproduce the BMS algebra, we will need the first $u$-moment of $\TTTten_\eps^{\vv3\vv\vv\vv\vv}.$ We use the same decomposition as in the main text and write  
\begin{align}
\TTTten_\eps^{\vv3\vv\vv\vv\vv}(C_T,b,c) = {}^1\TTTten_\eps^{\vv3\vv\vv\vv\vv}C_T+ {}^0\TTTten_\eps^{\vv3\vv\vv\vv\vv}(a,c)\, .\label{eq:split1}
\end{align}
The term proportional to $C_T$ in \eqref{eq:split1} will reproduce the BMS algebra in section \ref{sec:BMS}, but we find that the first moment of ${}^0\TTTten_\eps^{\vv3\vv\vv\vv\vv}$ is zero, i.e.
\begin{align}
\label{uintvA}
\lim_{\eps\to 0} \int du u \int d^2x_\perp  {}^0\TTTten^{\vv3\vv\vv\vv\vv}_{\eps}(u,x_\perp;w) =0\, ,
\end{align}
which will turn out to be necessary when computing various commutators of light-ray operators involved in the BMS algebra.

\subsection*{$\boldsymbol{[T^{\vv A}(u,0,x_\perp),T^{\vv B}(0)]|_T}$}

Setting $w=(0,w^\vv,0,w_3)$ we define $\TTTten_\eps^{\vv2\vv3\vv\vv}$ following (\ref{tttels}) and compute
\begin{align}
\TTTten^{[\vv 2,\vv 3]\vv 2}_\text{int.}(w) = \int d^2x_\perp du~ \<[T^{\vv 2}(u,0,x_\perp),T^{\vv 3}(0)] T^{\vv 2}(w)\>\, .
\end{align}
using (\ref{tttelsi}). We find
\begin{align}
\label{gv2v3v2}
\TTTten^{[\vv 2,\vv 3]\vv 2}_\text{int.}(w)= -C_T{8 i (w^\vv)^2 \over  (w_3)^{11}} = 2i  \p_{x_2}\<T^{\vv 3}(u,0,x_2,x_3)T^{\vv 2}(0,w^\vv,0,w_3)\>|_{u=x_2=x_3=0}\, ,
\end{align}
which implies
\begin{align}
\label{tuAuBapp}
[T_{u A}(u,0,x_\perp),T_{u B}(u',0,x'_\perp)]|_T& = i\left(T_{u B}(u,0,x_\perp)\p_A +T_{u A}(u',0,x'_\perp)\p_B\right)  \d(x_\perp-x_\perp')\d(u-u') + \text{td}\, .
\end{align}
We note that the result we find in (\ref{gv2v3v2}) is also equal to ${4i\over 5}  \p_{x_3}\<T^{\vv 2}(u,0,x_2,x_3)T^{\vv 2}(0,w^\vv,0,w_3)\>|_{u=x_2=x_3=0}.$ This would seem to imply a term $T_{u A}(x)\p_B$ in (\ref{tuAuBapp}), but upon integrating over $u$ and $x_\perp$ this term would lead to $[P_A,T_{uB}] \sim \p_B T_{uA}$ which is ruled out by Poincar\'e symmetry.

\section{Light scalars and supersymmetry}
\label{app:lssusy}

In this appendix we review the field content of $\cN=1$ superconformal field theories in $d=4$ dimensions and show that, in the absence of flavor symmetry, there does not exist R-symmetry neutral light scalars, ie. scalar conformal primaries with dimension $1\leq \D \leq 2.$ If such scalars existed they would have a chance to appear in the $TT$ OPE in the lightcone limit, and they would appear in the BMS algebra or the algebra of 4d Virasoro's we introduced in section \ref{sec:4dvirint}.

We follow \cite{Cordova:2016emh}. The $\cN=1$ superconformal algebra is $\mathfrak{su}(2,2|1)$ containing a $\mathfrak{u}(1)_R$ symmetry; the bosonic symmetry algebra is $\mathfrak{so}(4,2) \times \mathfrak{u}(1)_R$. In Table 12 in \cite{Cordova:2016emh} the bottom components of possible superconformal multiplets are listed in a unitary theory with notation $[j;\bar{j}]_\D^{(r)}.$ Expressing the complexified Lorentz algebra as $\mathfrak{so}(4)=\mathfrak{su}(2) \times \mathfrak{su}(2)$ the two integers $j \geq 0$ and $\bar{j} \geq 0$ specify $\mathfrak{su}(2)$ representations by their Dynkin labels. $r \in \mathbb{R}$ denotes a representation of the R symmetry, and $\D$ denotes the scaling dimension of the superconformal primary subject to unitarity bounds and shortening conditions. The $\mathfrak{su}(2)$ weights $j,\bar{j}$ specify a tensor sum of  $\mathfrak{so}(4)$ representations according to the Racah-Speiser algorithm $[j] \otimes [\bar{j}] = [|j-\bar{j}|] \oplus [|j-\bar{j}|+2] \oplus \ldots \oplus [j+ \bar{j}].$

Our search for light scalars in $\cN=1$ theories proceeds by picking representations in Table 12 that have dimension $1\leq \D \leq 2$ and looking for R-neutral light scalar conformal primaries that appear in these supermultiplets. The operator contents of these superconformal multiplets are obtained by repeated applications of the supercharge $Q$ on the superconformal primary and listed in section 4.5 of \cite{Cordova:2016emh}. We do not need to investigate the operator content of superconformal multiplets whose bottom operators have $\D >2$ since acting with $Q$ on such an operator we will never get a light scalar.

As an example we pick superconformal multiplets of type $\boldsymbol{A}_1 \boldsymbol{\overline{B}}_1.$ Their bottom components have quantum numbers $[j \geq 1;\bar{j}=0]_{\D={j\over 2}+1}^{(r={j+2\over 3})}.$ First of all, we notice this cannot be an R-neutral light scalar since that would require a negative value for $j$. Next we look at the superconformal descendants. The only other conformal primary in this multiplet is given as $[j +1;0]^{(r-1)}_{\D+{1\over 2}}.$ For this to be an R-neutral operator we need $r=1.$ This implies the bottom component has $j=1,$ so our descendant conformal primary is really of the form $[2;0]^{(0)}_{2}.$ However, this is a vector representation of the Lorentz algebra since $[2] \otimes [0] = [2].$

Another example is $\boldsymbol{L} \boldsymbol{\overline{B}}_1$ which has bottom component $[j;\bar{j}=0]_{\D={3r\over 2}}^{(r>{j+2\over 3})}.$ Using $j\geq 0$ this operator has $r>{2\over 3}$ and $\D>1;$ as such its Lorentz decomposition could contain a light scalar but not an R-neutral one. We turn to its superconformal descendants. These are $[j \pm 1;0]^{(r-1)}_{\D+{1\over 2}}$ and $[j ;0]^{(r-2)}_{\D+1}.$ The latter is heavy since $\D+1>2$ so we focus on the former. R-neutrality requires $r=1$ which puts our conformal primary into form $[j \pm 1;0]^{(0)}_{2}.$ This might potentially be a light scalar, depending on $j.$ To see if this is really allowed we need to go back to the bottom component which now looks like $[j;\bar{j}=0]_{\D={3\over 2}}^{(r=1>{j+2\over 3})}.$ This implies $j<1,$ which combined with $j\geq 0$ implies $j=0.$ Our descendant conformal primary has the final form $[\pm 1;0]^{(0)}_{2}.$ But according to the Racah-Speiser rules there is no Lorentz scalar here.

Another example that is worth discussing is $\boldsymbol{A}_2 \boldsymbol{\overline{A}}_2$ which has bottom component $[j=0;\bar{j}=0]_{\D=2}^{(r=0)}.$ This is an R-neutral light scalar, however, it comes with a flavor current in its superconformal multiplet as discussed in section 5.5.1 in \cite{Cordova:2016emh}. All the other superconformal multiplets in $\cN=1$ theories are free from R-neutral light scalars. In conclusion, in theories with $\cN=1$ superconformal symmetry and no flavor symmetry in $d=4$ there exist no R-neutral scalars with scaling dimension $1\leq \D \leq 2.$ This implies the stress tensor and the identity operator are the unique contributors to the lightsheet commutator of two stress tensors and our derivation of the BMS algebra is valid.

\section{BMS algebra computations}
\label{app:BMS}

In this appendix we finish the BMS algebra computations initiated in section \ref{subsec:BMS}.

\subsection*{$\boldsymbol{[\cK(x_\perp),\cK(x'_\perp)]}$}

We integrate (\ref{tuuuu}) to compute the commutator, which gets rid of the total derivative terms due to (\ref{uintvv}), leaving
\begin{align}
[\cK(x_\perp),\cK(x'_\perp)] &=i \d(x_\perp-x_\perp')\int_{-\infty}^{\infty}du \, u \int_{-\infty}^{\infty}du' \, u' \left(T_{uu}(u,0,x_\perp)+T_{uu}(u',0,x'_\perp)\right)\p_u \d(u-u')\no\\
&\equiv i\d(x_\perp-x_\perp')\cC_{\cK\cK}(x'_\perp)\, .
\end{align}
The first term contributes to $\cC_{\cK\cK}(x'_\perp)$ as
\begin{align}
-\int_{-\infty}^{\infty}du' \, u' T_{uu}(u',0,x_\perp) - \int_{-\infty}^{\infty}du' \, u'^2   \p_{u'} T_{uu}(u',0,x_\perp)\, ,
\end{align}
whereas the second gives
\begin{align}
\int_{-\infty}^{\infty}du \, u T_{uu}(u,0,x'_\perp) + \int_{-\infty}^{\infty}du \, u^2   \p_{u} T_{uu}(u,0,x'_\perp)\, .
\end{align}
Adding the two terms after suitably relabelling the integration variables and recalling the delta function in front we get 
\begin{align}
[\cK(x_\perp),\cK(x'_\perp)] =0\, .
\end{align}

\subsection*{$\boldsymbol{[\cN_{A}(x_\perp), \cK(x'_\perp)]}$}

We integrate (\ref{tuAuu}) to compute the commutator, which gets rid of the total derivative terms due to (\ref{uintvA}), leaving
\begin{align}
[\cN_{A}(x_\perp), \cK(x'_\perp)] &=i\int_{-\infty}^{\infty}du \int_{-\infty}^{\infty}du'u' T_{uu}(u,0,x_\perp) \d(u-u')\p_A \d(x_\perp-x_\perp')\no\\
&= -i \cK(x_\perp) \p_{A'} \d(x_\perp-x_\perp')\, .
\end{align}
We use the same trick as in the main text, namely add $i\p_{A'}\left[\d(x_\perp-x_\perp')\left( \cK(x_\perp)-\cK(x'_\perp) \right)\right]=0$ to express the operators at position $x'_\perp,$ 
\begin{align}
[\cN_{A}(x_\perp), \cK(x'_\perp)] &=-i \d(x_\perp-x_\perp') \p_{A'}\cK(x'_\perp) + i \cK(x'_\perp)\p_{A} \d(x_\perp-x_\perp')\, .
\end{align}

\subsection*{$\boldsymbol{[\cN_{A}(x_\perp), \cN_{B}(x'_\perp)]}$}

For this one we need (\ref{tuAuB})
\begin{align}
[T_{u A}(u,0,x_\perp),T_{u B}(u',0,x'_\perp)]|_T& = i\left(T_{u B}(u,0,x_\perp)\p_A +T_{u A}(u',0,x'_\perp)\p_B\right)  \d(x_\perp-x_\perp')\d(u-u')\, .
\end{align}
We thus compute the commutator as 
\begin{align}
[\cN_{A}(x_\perp), \cN_{B}(x'_\perp)] &=i\int_{-\infty}^{\infty}du \int_{-\infty}^{\infty}du' \left(T_{u B}(u,0,x_\perp)\p_A +T_{u A}(u',0,x'_\perp)\p_B\right)  \d(x_\perp-x_\perp')\d(u-u')\no\\
&= i \left( \cN_{B}(x_\perp) \p_{A} + \cN_{A}(x'_\perp) \p_{B}\right) \d(x_\perp-x_\perp')\, .
\end{align}
Adding to this $i\p_{A'}\left[\d(x_\perp-x_\perp')\left( \cN_{B}(x_\perp) - \cN_{B}(x'_\perp) \right)\right]=0$, we obtain finally
\begin{align}
[\cN_{A}(x_\perp), \cN_{B}(x'_\perp)] = -i \d(x_\perp-x_\perp')\p_{A'} \cN_{B}(x'_\perp) +i \cN_{B}(x'_\perp)\p_{A} \d(x_\perp-x_\perp') +i \cN_{A}(x'_\perp)\p_{B} \d(x_\perp-x_\perp')\, .
\end{align}
This concludes the computations of the generalized BMS algebra commutation relations.

\section{Details of 4d Virasoro}
\label{app:vir4}

In this appendix we report partial results for the commutators of the four-dimensional Virasoro operators introduced in (\ref{ceav}) in the main text. We compute the stress-tensor and identity contributions. As usual our strategy is to integrate the local commutator (\ref{tuuuu}).

\subsection{Identity contribution}
\label{sec:vir4id}

We compute the identity contribution to commutators of (\ref{ceav}), which we reprint here, 
\begin{align}
\label{eq:apLn}
\mathcal{L}_n(\vv, x_\perp)\equiv R^{-3}\int_{-\infty}^{\infty}du (iR+u)^{-n+2} (iR-u)^{n+2} T_{uu}(u,\vv,x_\perp)\, .
\end{align}
We recall for the central term (\ref{tvvvvi}), we obtained 
\begin{align}
{\<[T_{uu}(u,\vv,x_\perp),T_{uu}(0)]\>\over C_T} = -{i\pi\over 240\vv^2} \d^{(5)} \Big(u-{x_\perp^2 \over \vv}\Big)\, .
\end{align}
Stripping off a factor of $C_T$ we compute the identity contribution defined as
\be 
c_{m,n} = {\<[\mathcal{L}_m(\vv, x_\perp), \mathcal{L}_n(0)]\> \over C_T}\, .
\ee
which is given as 
\begin{align}
c_{m,n} &= -\frac{i\pi}{240 R^6 \vv^2}\int du'(iR + u')^{2-n}(iR-u')^{n+2}\int du (i R + u)^{2-m}(iR-u)^{m+2}\delta^{(5)}\left(u-u'-\frac{x_\perp^2}{\vv}\right)\, .
\end{align}
We perform the $u$ integral using the delta function, and then rename $u'\rightarrow u$. The result is 
\begin{align}
c_{m,n} &= {2\pi m(m^4-5m^2+4)\over 15 R \vv^2}\int_{-\infty}^\infty du(iR + u)^{2-n}(iR-u)^{n+2}\left(iR + u + \frac{x_\perp^2}{\vv}\right)^{-3-m}\left(iR - u - \frac{x_\perp^2}{\vv}\right)^{m-3}\, .
\end{align}
It is not hard to perform this integral for given $m$ and $n.$

We illustrate some general features and the light sheet limit of $c_{m,n}.$ Any non-zero $c_{m,n}$ turns out to be a function of the combination $R\vv,$ $c_{m,n}=c_{m,n}(R\vv,x_\perp).$ By direct evaluation we determine non-zero values of $c_{m,n}$ and summarize this in Table \ref{table1}. 
\begin{table}[h!]
	\centering
	\begin{tabular}{|c|c|c|c|}
		\hline
		$c_{m,n}$       & $m\leq -3$ & $-2\leq m \leq 2$ & $m\geq 3$  \\\hline
		$ n\leq -3$            &  $0$       &     $0$    &     $\neq 0$  \\ \hline
		$-2\leq n \leq 2$    &     $0$    &     $0$    &   $0$   \\ \hline
		$n \geq 3$           &      $\neq 0$   &    $0$     &    $0$   \\ \hline
	\end{tabular}
	\caption{Values of $c_{m,n}$ for a given $m$ and $n$.}
	\label{table1}
\end{table}
The zeroes in Table \ref{table1} are a consequence of 
\begin{align}
\mathcal{L}_m\ket{0} &= 0\, , \qquad n\geq -2\, ,\\
\mathcal{L}_m^\dagger &= \mathcal{L}_{-m}\, .
\end{align}
$c_{m,n}$ satisfies the antisymmetry property $c_{m,n}(R\vv,x_\perp)=-c_{n,m}(-R\vv,-x_\perp)$ and vanishes in the light sheet limit, $\vv \to 0,$
for finite transverse separation,
\begin{align}
\label{cmnv0}
c_{m,n}(0, x_\perp) = 0\, .
\end{align}
In the coincident transverse point limit it takes the form
\begin{align}
\label{cmnx0}
c_{m,n}(R\vv, 0)=-{2\pi^2\over 15 R^2 \vv^2}m(m^4-5m^2+4)\d_{m+n,0}\, .
\end{align} 
The relation (\ref{cmnx0}) combined with (\ref{cmnv0}) implies 
\begin{align}
\lim_{R\vv \to 0} c_{m,n}(R\vv, x_\perp) = \b_{m,n}\d(x_\perp)\, .
\end{align}
for some possibly divergent differential operators $\b_{m,n}$ acting on the tranverse delta function $\d(x_\perp)=\d(x_2)\d(x_3)$. For example
\begin{align}
c_{3,-3}(R\vv, x_\perp) &= -1024 \pi^2 {(R\vv)^4 \over (2R\vv-i x_\perp^2)^6}\, .
\end{align} 
By dimensional analysis this expression is a linear combination of $\lim_{R\vv \to 0}{1\over R\vv}\d(x_\perp)$ and $\D_\perp\d(x_\perp)$ in the light sheet limit $R\vv \to 0,$ where $\D_\perp =\p_2^2+\p_3^2$ denotes the transverse Laplacian. We can determine the coefficients of these two structures by integrating $c_{3,-3}(R\vv, x_\perp)$ against suitable test functions. In fact, we compute in full generality,
\begin{align}
\lim_{R\vv \to 0} c_{m,n}(R\vv , x_\perp) = {2\pi^2\over 75 }|m|(m^4-5m^2+4)\d_{m+n,0} \left({m\pi \over 12}\D_\perp\d(x_\perp) - \lim_{R\vv  \to 0} \left[ {2i\pi \over R\vv } \d(x_\perp) \right]\right) \, .
\end{align}

\subsection{Stress-tensor contribution}
\label{app:vir4dst}

Without a proper analysis of the total derivative terms in (\ref{tuuuu}), which contain theory specific information and are outside the scope of this work, we cannot make a proper analysis of the stress-tensor contribution to the commutators of $\mathcal{L}_m$ operators. The goal of this appendix is to investigate what we obtain if we neglect all total derivative terms. 

We define a general class of integrals of the stress-tensor as
\begin{align}
\label{l4dgen}
\mathcal{L}_n^{\text{gen}}(x_\perp)=R^{1-\a-\b-\g}\int_{-\infty}^{\infty}du (iR+u)^{-n+\a} (iR-u)^{n+\b} (s-u)^\g T_{uu}(u,0,x_\perp)\, .
\end{align}
where $R,s \in \mathbb{R}_{>0}.$ These operators have mass dimension 2, satisfy $(\mathcal{L}_n^{\text{gen}})^\dagger =\mathcal{L}_{-n}^\text{gen}$ and have finite correlators for $\a+\b+\g=4.$ Ignoring the total derivative terms in (\ref{tuuuu}), we obtain 
\begin{align}
[\mathcal{L}_m^\text{gen}(x_\perp), \mathcal{L}_n^\text{gen}(x'_\perp)]|_T &\equiv \d^{(d-2)} (x_\perp-x'_\perp) f_{m,n}(x_\perp)\, ,
\end{align}
where
\begin{align}
f_{m,n}(x_\perp) &=2iR^{3-2(\a+\b+\g)}(m-n)  \int_{-\infty}^{\infty}du~(iR+u)^{-m-n+2\a-1} (iR-u)^{m+n+2\b-1} (s-u)^{2\g} T_{uu}(u,0,x_\perp)\, .
\end{align}
Namely if we set $\a=\b=2,~\g=0$, which is how we defined $\mathcal{L}_n(x_\perp)$ in \eqref{ceav}, we obtain
\begin{align}
f_{m,n}(x_\perp) &=-2iR^{-5}(m-n)  \int_{-\infty}^{\infty}du~(iR+u)^{-m-n+2} (iR-u)^{m+n+2} (R^2+u^2) T_{uu}(u,0,x_\perp)\, .
\end{align}
For $\a=\b=1,$ and $\gamma=0$ a nicer structure emerges, with 
\begin{align}
f_{m,n}(x_\perp) &=2iR^{-1}(m-n)  \int_{-\infty}^{\infty}du~(iR+u)^{-m-n+1} (iR-u)^{m+n+1} T_{uu}(u,0,x_\perp)\\
&=2i(m-n)\cL_{m+n}(x_\perp)\, .
\end{align}
A commutator of this form was conjectured in \cite{Casini:2017vbe}. We have shown how this can be obtained from local stress-tensor commutators. Let us reiterate that to put this claim on firm ground, a careful analysis of the total derivative terms is needed. This is beyond the scope of this work.

\section{Integrals of the stress-tensor two-point function}
\label{app:itt}

Poincar\'e charges, obtained by spatial integrals of the stress-tensor,
\begin{align}
P^\m = \int d^3x T^{0\m}(t,\vec{x})\, , 
\end{align}
annihilate the vacuum. We verify this in the example of the stress tensor two-point function. Recalling the shorthand (\ref{gx})
\begin{align}
\TTten^{\m\n\s\r}(x) = {\< T^{\mu \nu}(x) T^{\sigma \rho}(0)\>\over C_T}\, ,
\end{align}
we obtain ordered correlation functions by using the $i\eps$ prescription as
\begin{align}
{\<T^{00}(0,\vec{x})T^{00}(0)\>\over C_T}  = \lim_{\eps\to 0} \TTten^{0000}(-i\eps,\vec{x})= \lim_{\eps\to 0} \left[ {3\over 4(r^2+\eps^2)^4} - {4\eps^2\over (r^2+\eps^2)^5}  + {4\eps^4\over (r^2+\eps^2)^6}\right]\, .
\end{align}
The spatial integral of this expression vanishes
\begin{align}
\int d^3x \TTten^{0000}(-i\eps,\vec{x}) =0\, .
\end{align}
Similarly $\int d^3 x \TTten^{\m\n\s\r}(-i\eps,\vec{x})=0$ for $(\m\n\s\r)=(000i),(00ij),(0ijk)$.

We can do the same computation in lightcone coordinates where we use 
\begin{align}
{\<T^{\m\n}(u-i\eps,\vv-i\eps,x_2,x_3)T^{\s\r}(0)\>\over C_T}  = \lim_{\eps\to 0} \TTten^{\m\n\s\r}(u-i\eps,\vv-i\eps,x_2,x_3)\, .
\end{align}
We find $\int_{-\infty}^\infty du~ \TTten^{\m\n\s\r}(u-i\eps,\vv-i\eps,x_2,x_3) =0$ for $(\m\n\s\r)=(\vv\vv \s\r),(\vv A \s\r)$ and $\int_{-\infty}^\infty du\, u~ \TTten^{\vv\vv \s\r}(u-i\eps,\vv-i\eps,x_2,x_3) =0$ for arbitrary $\s,\r.$ These imply
\begin{align}
\<\cA(\vv,x_\perp) T^{\m\n}(0)\>=0\, .
\end{align} 
for $\cA=\cE,\cK,\cN_A.$ For our Virasoro operators (\ref{ceav})
\begin{align}
\<\cL_n(\vv,x_\perp) T^{\m\n}(0)\>=0,~~~n\leq 2.
\end{align}

\section{Example of equal-time stress-tensor commutator}
\label{app:etete}

To gain further confidence in our technique in this appendix we compute the stress-tensor contribution to $[T^{0k}(\vec{x}),T^{0m}(0)].$ We will verify
\begin{align}
\label{okome}
i[T^{0k}(\vec{x}), T^{0m}(\vec{y})] &= \left(T^{0m}(\vec{x})\p^k+T^{0k}(\vec{y})\p^m\right) \d(\vec{x}-\vec{y}) +\text{td}\, .
\end{align}
which is implied by Poincar\'e symmetry. We compute
\begin{align}
\TTTten^{010202}(x;w)&=\< T^{01}(x) T^{02}(0) T^{02}(w)\>\\
\TTTten^{[01,02]02}_\text{int.}(w) &=\int \<[T^{01}(0,\vec{x}), T^{02}(0)] T^{02}(w)\> d^3x
\end{align}
via
\begin{align}
\TTTten^{010202}_{\eps}(\vec{x};w) & \equiv \TTTten^{010202}(-i \eps,\vec{x};w) -\TTTten^{010202}(i\eps,\vec{x};w),\\
\TTTten^{[01,02]02}_\text{int.}(w) &= \lim_{\eps\to 0} \int d^3x~ \TTTten^{010202}_{\eps}(\vec{x};w)\, .
\end{align}
As in the main text we simplify $\TTTten^{010202}_{\eps}(\vec{x};w)$ by setting $w=(w^0,w_1,0,0).$ To manage the spatial integral we expand $\TTTten^{010202}_{\eps}(\vec{x};w)$ in powers of $x^2=r^2+\eps^2$ as
\begin{align}
\TTTten^{010202}_{\eps}(\vec{x};w)={\TTTten^{010202}_{\eps,5}(\vec{x};w)\over (r^2+\eps^2)^5} + {\TTTten^{010202}_{\eps,4}(\vec{x};w)\over (r^2+\eps^2)^4} + {\TTTten^{010202}_{\eps,3}(\vec{x};w)\over (r^2+\eps^2)^3} + {\TTTten^{010202}_{\eps,2}(\vec{x};w)\over (r^2+\eps^2)^2} \, ,
\end{align}
where we have written only the terms that do not vanish in the $\eps \to 0$ limit as distributions in $\vec{x}.$ Excluding a term in $\TTTten^{010202}_{\eps,2}(\vec{x};w)$ that gives a divergent integral which we treat separately below, the rest of the terms combine to give
\begin{align}
\label{g010202}
\TTTten^{[01,02]02}_\text{int.}(w) = -2i C_T {w_1\left[3(w^0)^2 + 2(w_1)^2\right]\over (-(w^0)^2+(w_1)^2)^6}\, .
\end{align}
This is equal to $i\p_1 \<T^{02}(0,x_1,x_2,x_3)T^{02}(w^0,w_1,0,0)\>|_{x_1=x_2=x_3=0}.$ implying the relation (\ref{okome}).

We turn to the divergent integral coming from $\TTTten^{010202}_{\eps,2}(\vec{x};w).$ We first perform the $x_3$ integral
\begin{align}
\label{ix1x2}
I(x_1,x_2) \equiv \int_{-\infty}^\infty dx_3 {\TTTten^{010202}_{\eps,2}(\vec{x};w)\over (x_1^2+x_2^2+x_3^2+\eps^2)^2} = {N(x_1,x_2) \over w^8 (x_1^2+x_2^2+\eps^2)^{3/2}(w^4-4w^2(w_1 x_1+\eps^2)+4w_1^2(x_1^2+\eps^2))^6}\, ,
\end{align}
where $N(x_1,x_2)$ is of the form
\begin{align}
N(x_1,x_2) = C_0 + C_2 x_2^2\, ,
\end{align}
where $C_0$ and $C_2$ are complicated coefficients that depend on $x_1,\eps,w.$ The $O(x_2^0)$ term is harmless, we perform its $x_1$ and $x_2$ integrals and this goes into the result (\ref{g010202}). The $x_2$ integral of the $O(x_2^2)$ term diverges. However this term is actually equal to zero as a distribution in $x_1,x_2$ in the $\eps \to 0$ limit. To see this we denote the denominator in (\ref{ix1x2})  $D(x_1,x_2)$ and observe
\begin{align}
\lim_{\eps \to 0}  {x_2^2 C_2(x_1,\eps,w)\over D(x_1,x_2)} = 0,~~~~\lim_{x_1,x_2 \to 0}  {x_2^2 C_2(x_1,\eps,w)\over D(x_1,x_2)} = 0.
\end{align}

\bibliographystyle{ytphys}
\bibliography{ref}

\end{document}